# Analytical modeling of micelle growth. 4. Molecular thermodynamics of wormlike micelles from ionic surfactants: theory vs. experiment

Krassimir D. Danov [a], Peter A. Kralchevsky [a,*], Rumyana D. Stanimirova [a], Simeon D. Stoyanov [b,c,d], Joanne L. Cook [e], Ian P. Stott [e]

[a] *Department of Chemical and Pharmaceutical Engineering, Faculty of Chemistry and Pharmacy, Sofia University, Sofia 1164, Bulgaria*

[b] *Unilever Research & Development Vlaardingen, 3133AT Vlaardingen, The Netherlands*

[c] *Laboratory of Physical Chemistry and Colloid Science, Wageningen University, 6703 HB Wageningen, The Netherlands*

[d] *Department of Mechanical Engineering, University College London, WC1E 7JE, UK*

[e] *Unilever Research & Development Port Sunlight, Bebington CH63 3JW, UK*

ORCID Identifiers: Krassimir D. Danov: 0000-0002-9563-0974 ; Peter A. Kralchevsky: 0000-0003-3942-1411 ; Simeon D. Stoyanov: 0000-0002-0610-3110

ABSTRACT

*Hypotheses:* The aggregation number and length of spherocylindrical (rodlike, wormlike) micelles in solutions of an ionic surfactant and salt can be predicted knowing the molecular parameters and the input concentrations of the species. This can be achieved by upgrading the quantitative molecular thermodynamic model from the previous parts of this series with an expression for the electrostatic component of micelle scission energy that is the excess free energy of the spherical endcaps with respect to the cylindrical part of the micelle.

*Theory:* The thermodynamics of micellization is extended to the case of multicomponent system, which may contain several surfactants (both ionic and nonionic) and salts, taking into account the effect of counterion binding in the Stern layer on the micellar surface. Furthermore, the considerations are focused on a system that consists of single ionic surfactant plus salt.

*Findings:* Excellent agreement was achieved between the theoretical model and experimental data for wormlike micelles from anionic and cationic surfactants at various concentrations of salt and temperatures. In accord with the experimental observations, at high salt concentrations, the model predicts loss of chemical equilibrium between the endcaps and cylindrical part of the wormlike micelles, which implies transition to self-assemblies of other, e.g. branched, morphology or the onset of crystallization and phase separation.

* Corresponding author. Tel.: +359 2 962 5310
  E-mail address: pk@lcpe.uni-sofia.bg (P.A. Kralchevsky)

# 1. Introduction

In the previous parts of this series, the molecular thermodynamic theory of wormlike micelles was upgraded to a form, which provides excellent agreement with experimental data for the mean mass aggregation number, $n_M$, of micelles from nonionic surfactants [1] and from nonionic surfactant mixtures [2,3]. The next step has been to extend this approach to *ionic* wormlike micelles in the presence of salt. For this goal, in Ref. [4] we addressed the issue about the accurate calculation of the electrostatic component of micelle free energy taking into account the effect of mutual spatial confinement of the electric double layers (EDLs) of the neighboring micelles, as well as the effect of ionic activity coefficients. Here, the electrostatic free energy component, calculated according to [4], is combined with the other micelle free energy components to predict the micellar size and the results are compared with available experimental data for the size of wormlike micelles from both anionic and cationic surfactants at various salt concentrations and temperatures.

In micellar solutions containing ionic surfactants, the dependence of viscosity on the concentration of added salt (the so called salt curve) often exhibits a high peak [5-11]. This peak could be interpreted as a transition from wormlike micelles to branched micelles [8,12-17]. The initial growth of wormlike micelles could be explained with the screening of the electrostatic repulsion between the surfactant headgroups by the electrolyte, whereas the subsequent transition to branched aggregates can be interpreted in terms of surfactant packing parameters and interfacial bending energy [18]. At high salt concentrations, the viscosity could drop because of phase separation due to the salting out of surfactant [19].

A basic review on micellization, including information on the second critical micellization concentration (2$^{nd}$ CMC); the persistence length of wormlike micelles, and the stability of linear micelles was published by Leermakers et al. [20]. A recent review on the theory of wormlike micelles (WLM) can be found in Ref. [1]. For this reason here we focus on papers closer related to the present study. The majority of papers in this field is aimed at predicting the rheological behavior of micellar solutions at given micelle concentration, size distribution and kinetic parameters, such as the characteristic times of micelle breakage and reptation [21-23]. There are also studies on computer simulations of wormlike micelles by the Monte Carlo [24,25] and molecular dynamics [26,27] methods. The application of simulation methods is limited because of the large size of the wormlike micelles – typically, their aggregation number ranges from thousand to million molecules. Paths toward overcoming of these size limitations is the method of dissipative particle dynamics (DPD) [28-30], and the recently proposed density field thermodynamic integration approach [31]. Here we apply another method, the molecular thermodynamic theory, which does not face any micelle-size limitations, and which is able (in principle) to predict micelle aggregation number and length, based on information about the molecular parameters and available experimental data for some collective properties, such as interfacial tension and activity coefficients.

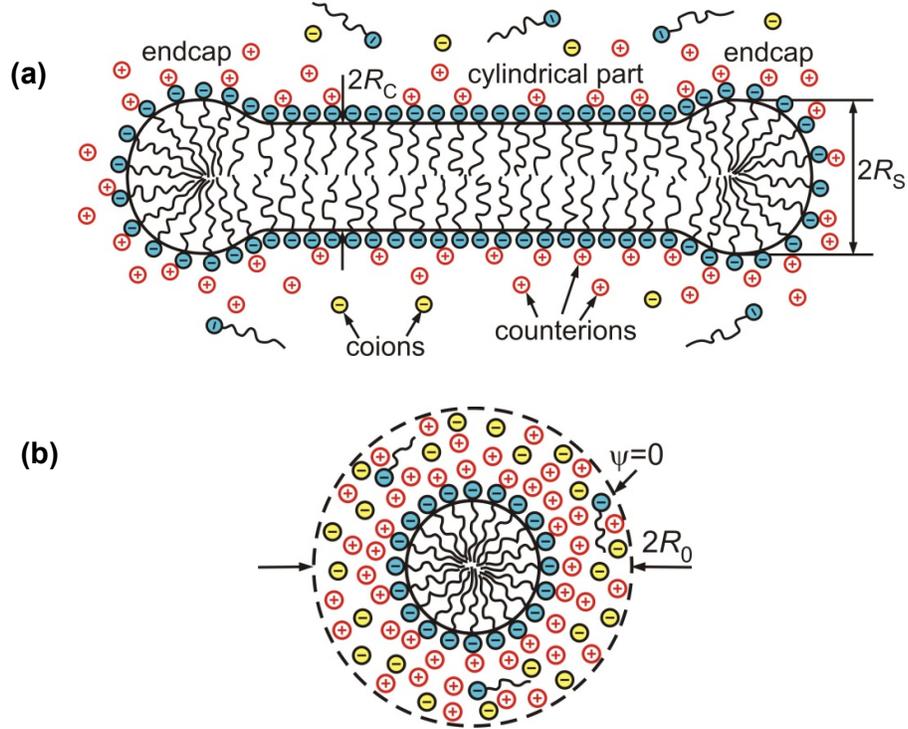

**Fig. 1.** (a) Sketch of a spherocylindrical micelle from anionic surfactant; $R_c$ and $R_s$ are the radii of the micellar hydrocarbon core in the regions of the cylindrical part and the endcaps. (b) Cross-section of a cylindrical (or spherical) micelle; the used cell model [4] assumes that the electric double layer around the micelle is closed in a cell of outer radius $R_0$, at which the electric potential is set zero, $\psi = 0$.

The cryo-transmission electron microscopy confirms that one of the basic self-assemblies formed in solutions of ionic surfactants in the presence of salt are the spherocylindrical (rodlike, wormlike) micelles (Fig. 1a) [16,17,32-34]. Missel et al. [35] established that the mean mass aggregation number of such micelles can be estimated from the expression:

$$n_M = 2[(X_S - X_S^o)\exp(E_{sc})]^{1/2} \qquad (1.1)$$

where $X_S$ is the total molar fraction of surfactant in the aqueous solution; $X_S^o$ can be identified (in first approximation) with the solution's critical micellization concentration (CMC); $E_{sc}$ is the excess free energy (in $k_BT$ units) of the molecules in the two spherical micelle endcaps with respect to the cylindrical part of the micelle ($k_B$ is the Boltzmann constant, and $T$ is the temperature). Because the breakage (scission) of a long spherocylindrical micelle to two smaller micelles leads to the formation of two new endcaps, $E_{sc}$ was termed *scission energy* in relation of the reptation-reaction model of the rheology of micellar solutions developed by Cates and coworkers [22,23]. For not-too-short spherocylindrical micelles (negligible edge-overlap effects), $E_{sc}$ is independent of the surfactant concentration, $X_S$. Hence, knowing $E_{sc}$ one can predict the micelle aggregation number, $n_M$, for each given $X_S$ using Eq. (1.1). Furthermore, from the obtained $n_M$, one can estimate micelle length, $L$:

$$L \approx \frac{n_M}{\pi R_c^2} \sum_{j=1}^{m} y_j v_j \tag{1.2}$$

Here, for a long micelle it is (approximately) assumed that the whole micelle is a cylinder; $R_c$ is the radius of the micellar hydrocarbon core (Fig. 1a); $m$ is the number of surfactant species in the micelle; $y_j$ are their molar fractions and $v_j$ are the volumes of their hydrophobic chains.

Thus, the main goal of the molecular thermodynamic theory is the prediction of $E_{sc}$. For single-component spherocylindrical micelles, $E_{sc}$ can be expressed in the form [1,35-37]:

$$E_{sc} = n_s (f_s - f_c) / (k_B T) \tag{1.3}$$

where $n_s$ is the total aggregation number of the two micelle endcaps (with shapes of truncated spheres; Fig. 1a); $f_s$ and $f_c$ are the free energies per molecule in the endcaps and in the cylindrical part of the micelle, respectively. Typically, $E_{sc}$ varies in the range 15–30 (see below) and $n_s$ – in the range 60–120. Consequently, the difference $(f_s - f_c)$ varies in the range 0.125–0.5 $k_BT$. In other words, to predict correctly $n_M$ and $L$, the molecular thermodynamic theory has to predict $(f_s - f_c)$ within accuracy better than 0.01 $k_BT$, i.e., $f_s$ and $f_c$ have to be very accurately predicted. This is the main difficulty with the molecular thermodynamics of micelle growth. In Refs. [1,3], we have demonstrated that this difficulty can be overcome for micelles from nonionic surfactants.

In the case of wormlike micelles from *ionic* surfactants, there are only few molecular-thermodynamic studies, where the theoretically predicted $n_M$ is compared with experimental data for $n_M$. In the papers by Missel et al. [35,38,39] for sodium alkyl sulfates at relatively high NaCl concentrations, the validity of Eq. (1.1) for $n_M$ vs. $X_S$ is demonstrated; the micelle growth parameter $K = \exp E_{sc}$ is determined from experimental plots and the obtained $E_{sc}$ values are interpreted in the framework of a semiempirical model. Results are presented for the growth of ionic wormlike micelles with the rise of both salt concentration and temperature [38,39].

In their basic paper on micellar growth, Nagarajan and Ruckenstein [36] employed formula derived by Ninham et al. [40,41] to estimate the electrostatic component of micelle free energy. The effect of counterion binding was not taken into account. The comparison of the calculated $E_{sc}$ vs. $T$ curves with experimental data for sodium dodecyl sulfate (SDS) was not satisfactory [36].

The effect of counterion binding at micelle surface has been first taken into account by Alargova et al. [42,43] to explain the enhancement of micelle growth in the presence of $Ca^{2+}$ and $Al^{3+}$ counterions in solutions of an anionic surfactant. Next, this approach was extended to other counterions and surfactants by Srinivasan and Blankschtein [44]. Koroleva and Victorov [45] took into account the effect of the finite size of the ions in the electric double layer (EDL) around the micelle by using the Boublik−Mansoori−Carnahan−Starling−Leland (BMCSL) equation [46-48] for a mixture of hard spheres of different radii. The excluded volume of the hydrated ions becomes essential at high micelle concentrations and near the charged surfaces, similarly to the case of the thin liquid films [49].

Despite of these improvements of theory, a systematic agreement of the molecular-thermodynamic theory with the available sets of data for the aggregation number $n_M$ of ionic wormlike surfactant micelle has not been achieved in the preceding studies. As mentioned above, to achieve agreement between theory and experiment with respect to $n_M$, one has to accurately take into account all significant contributions to the micelle free energy in order to predict the difference $(f_s - f_c)$ within accuracy better than 0.01 $k_B T$. Such upgrade of the molecular-thermodynamic model and achievement of agreement between theory and experiment, i.e. quantitative prediction of the scission energy $E_{sc}$ and the aggregation number $n_M$ for ionic wormlike micelles, is our goal in the present study. Because of its large volume, our study has been split to two parts.

The previous part, Ref. [4], the approximate assumption that micelle electrostatic potential $\psi$ in a concentrated micellar solution decays at infinity [36,40-45] was avoided. Instead, the electric field was calculated using a cell model, which takes into account the mutual confinement of the EDLs of the neighboring micelles. The boundary condition $\psi = 0$ at $r \to \infty$ was replaced with a boundary condition at the outer cell border, $\psi = 0$ at $r = R_0$ (Fig. 1b). Very often, the micelle growth happens at high salt concentrations (0.5 – 4 M). For this reason, the effect of activity coefficients, $\gamma_j$, of the counterions and coions becomes important. In our study, theoretical expressions for $\gamma_j$ were used, which take into account (i) electrostatic; (ii) hard-sphere and (iii) specific interactions between the ions, and exactly describe the experimental dependencies of $\gamma_\pm = (\gamma_+\gamma_-)^{1/2}$ on the salt concentration; $\gamma_j$ varies across the micellar EDL as a function of the local ionic concentrations. A new, convenient and numerically precise approach to the calculation of electrostatic free energy was developed, which allows using one-time numerical solution of the differential-equation problem, instead of numerous solutions of this problem at each step of the numerical integration. At that, the effect of micelle surface curvature on the EDL was taken into account exactly (without using any truncated series expansions), working with variable ionic activity coefficients, $\gamma_j$, across the EDL [4]. The effect of counterion binding was taken into account via the Stern isotherm [50]. Such detailed description of the electrostatic effects with ionic surfactant micelles has been given in none of the preceding studies [36,40-45].

The present paper continues our study on wormlike micelles from *ionic* surfactants [4]. Section 2 is dedicated to the molecular thermodynamics of electrically charged micelles in a general form taking into account mixed micelles of ionic and nonionic surfactants in the presence of added electrolytes. Next, the theory is applied to the case of single ionic surfactant and an explicit expression for $E_{sc}$ is derived with account for the effect of counterion binding. Section 3 is focused on the effect of salt concentration on the oil/water interfacial tension. This effect is important for the micelle growth because among the four different components of $E_{sc}$, the interfacial tension component $(E_{sc})_\sigma$ turns out to be the greatest one. Section 4 describes the molecular model for calculation of all four components of $E_{sc}$, with an emphasis on the calculation of the electrostatic component, $(E_{sc})_{el}$. Furthermore, the predictions of the theoretical model are compared with experimental data for sodium dodecyl sulfate (SDS; Section 5); for sodium unidecyl and tridecyl sulfates (SUS and STS; Section 6), and for cationic surfactants, including cetyltrimethylammonium bromide and chloride (CTAB and CTAC; Section 7). The results and relations to preceding studies are discussed in Section 8. We could mention in advance that excellent agreement theory/experiment has been achieved with the variation of zero or (maximum) one adjustable parameter.

## 2. Molecular thermodynamics of electrically charged micelles

*2.1. General thermodynamic relations*

Here, our goal is to expand the molecular thermodynamics of mixed micellar surfactant solutions in Ref. [3] for the case of solutions containing ionic surfactants and electrolytes (salts). The final goal is to derive an expression for the micelle scission energy, $E_{sc}$, which would allow one to predict the mean aggregation number of the equilibrium wormlike/ rodlike micelles formed in such solutions. In comparison with the case of nonionic micelles, for charged micelles additional contributions to $E_{sc}$ originate from the diffuse electric double layer (EDL) and from the counterion binding to the surfactant headgroups on the micellar surface. Even for a micellar solution of single ionic surfactant, the system is multicomponent, because it contains surfactant ions; their counterions, and coions of added salt (if any).

Let us first consider the general case of micellar solution, which contains ionic and nonionic surfactants, and several electrolytes. The micelles are polydisperse in size and are composed of surfactant molecules and bound counterions. If the micelle contains both anionic and cationic surfactants, all cations and anions originating from dissolved salts play the role of counterions. In such mixed solution, the mass action law for a micellar aggregate reads:

$$g_k^o + k_B T \ln X_k = \sum_{j=1}^{m+m_C} k_j \mu_{1,j}; \quad k \equiv \sum_{j=1}^{m} k_j \qquad (2.1)$$

Here, $k$ is the total number of surfactant molecules in the micellar aggregate (the aggregation number); $g_k^o$ is the standard free energy of such $k$–aggregate; $X_k$ is the molar fraction of the $k$-aggregates in the aqueous solution; $k_B$ is the Boltzmann; $T$ is the temperature; $k_j$ for $1 \leq j \leq m$ is the number of type $j$ *surfactant* molecules in the micelle; $k_j$ for $m + 1 \leq j \leq m + m_C$ is the number of type $j$ bound *counterions* in the micelle; $m$ and $m_C$ are the numbers, respectively, of surfactant and counterion species in the solution; $\mu_{1,j}$ is the electrochemical potential of the type $j$ species in the form of free monomers in the solution. For a nonionic species, the electrochemical potential coincides with its chemical potential. In the left-hand side of Eq. (2.1), the effect of electrostatic interactions is included in $g_k^o$ (see below).

At this stage, we are dealing with micelles of arbitrary shape (spherical; spherocylindrical; discoidal, etc.); the shape of the micelles will be specified in Section 2.2. The electrostatic energy of the micellar aggregates will be calculated in the framework of a model [4] assuming that each aggregate is closed in a cell, which contains the aggregate and the adjacent diffuse electric double layer. The shape of the cell corresponds to the shape of the aggregate; see, e.g., Fig. 1b. Each cell is electroneutral. At its outer boundary (at $r = R_0$), the electric field is equal to zero, $\mathbf{E} = \mathbf{0}$, and the electrostatic potential can be also set equal to zero, $\psi = 0$, because the potential is defined up to an additive constant. In such a case, the electrochemical potential of the free monomers from all surfactant species (both ionic and nonionic) at the outer boundary of the cell can be written in the form

$$\mu_{1,j} = \hat{\mu}^o_{1,j} + k_B T \ln X_{1,j}, \quad j = 1, 2, ..., m \tag{2.2}$$

where $X_{1,j}$ is the molar fraction of the type $j$ free monomers in the aqueous phase at the outer cell boundary, i.e. at $r = R_0$ (Fig. 1b), and

$$\hat{\mu}^o_{1,j} \equiv \mu^o_{1,j} + k_B T \ln \gamma_{j,0}, \quad j = 1, 2, ..., m \tag{2.3}$$

where $\mu^o_{1,j}$ is standard chemical potential and $\gamma_{j,0}$ is activity coefficient. Of course, a relation in the form of Eq. (2.2) can be written also for all counterions and coions at the cell boundary. In view of further applications, for the counterions it is convenient to use the relation:

$$\mu_{1,j} = \mu_{a,j} \quad \text{for } j = m+1, 2, ..., m+m_C \tag{2.4}$$

where $\mu_{a,j}$ is the *electrochemical* potential of a counterion, which is bound to (adsorbed at) the surface of the micellar aggregate, i.e. counterion in the Stern layer. Eq. (2.3) is a corollary from the condition for electrochemical equilibrium in the EDL. It is convenient to introduce the notations:

$$x_j \equiv \frac{X_{1,j}}{X_1} \quad (j = 1, 2, ..., m) \tag{2.5}$$

$$X_1 = \sum_{j=1}^{m} X_{1,j}; \quad \sum_{j=1}^{m} x_j = 1 \tag{2.6}$$

Here, $X_1$ is the total molar fraction of surfactant monomers in the aqueous phase at the cell periphery and $x_j$ is molar fraction of the type $j$ surfactant in the surfactant blend. In view of Eqs. (2.2), (2.4) and (2.5), the mass action law, Eq. (2.1), can be represented in the form:

$$\ln X_k = \ln X_1^k - \frac{1}{k_B T} \left( g_k^o - \sum_{j=1}^{m} k_j (\hat{\mu}^o_{1,j} + k_B T \ln x_j) - \sum_{j=m+1}^{m+m_C} k_j \mu_{a,j} \right) \tag{2.7}$$

Taking inverse logarithm of the above equation, we obtain:

$$X_k = X_1^k \exp\left(-\frac{\Phi}{k_B T}\right) \tag{2.8}$$

where

$$\Phi = g_k^o - \sum_{j=1}^{m} k_j (\hat{\mu}_{1,j}^o + k_B T \ln x_j) - \sum_{j=m+1}^{m+m_C} k_j \mu_{a,j} \qquad (2.9)$$

Eqs. (2.8) and (2.9) describe the size distribution of charged micellar aggregates. It can be applied to aggregates of various shapes. The next step is to specify the shape of the aggregates.

*2.2. Size distribution of spherocylindrical micelles*

Let us consider spherocylindrical micelles, like that depicted in Fig. 1a, for which the micelle aggregation number can be presented in the form:

$$k = n_c + n_s; \quad n_c = \sum_{j=1}^{m} k_{c,j}; \quad n_s = \sum_{j=1}^{m} k_{s,j} \qquad (2.10)$$

where $n_c$ is the number of surfactant molecules in the cylindrical part of the micelle, whereas $n_s$ is the number of surfactant molecules in the two spherical endcaps together. (The endcaps have the shape of truncated spheres; see Fig. 1a.) Here and hereafter, the subscripts "c" and "s" refer, respectively, to the micelle cylindrical part and to the spherical endcaps. Likewise, the quantity $\Phi$ in Eq. (2.9) can be presented as a sum of contributions from the cylindrical part and from the endcaps:

$$\Phi = n_c f_c - n_c \sum_{j=1}^{m} y_{c,j} (\hat{\mu}_{1,j}^o + k_B T \ln x_j)$$
$$+ n_s f_s - n_s \sum_{j=1}^{m} y_{s,j} (\hat{\mu}_{1,j}^o + k_B T \ln x_j) \qquad (2.11)$$

where by definition

$$n_c f_c \equiv g_{c,k}^o - \sum_{j=m+1}^{m+m_C} k_{c,j} \mu_{a,j}; \quad n_s f_s \equiv g_{s,k}^o - \sum_{j=m+1}^{m+m_C} k_{s,j} \mu_{a,j}; \qquad (2.12)$$

$$y_{c,j} = \frac{k_{c,j}}{n_c}; \quad y_{s,j} = \frac{k_{s,j}}{n_s} \quad (j=1,...,k) \qquad (2.13)$$

Here $f_c$ and $f_s$ are the free energies per surfactant molecule in the cylindrical part and in the spherical endcaps; $y_{c,j}$ and $y_{s,j}$ are the surfactant molar fractions in the respective parts of the micelle. Eq. (2.12) indicates that in order to get the free energy per surfactant molecule, one has to subtract the Gibbs free energy of the bound counterions, $\Sigma k_j \mu_{a,j}$, from the total micelle standard free energy $g_k^o$.

For sufficiently long micelles ($n_c \gg n_s$), the local properties in the cylindrical part of the micelle become independent on its total aggregation number, $k$; see, e.g., Ref. [3]. Because the endcaps are in chemical equilibrium with the cylindrical part, their properties are also independent of $k$. Then, the micelle free energy, $\Phi$ in Eq. (2.11), becomes a linear function of $k$:

$$\Phi = Ck + E_{sc} k_B T \qquad (2.14)$$

where the slope $C$ and the intercept $E_{sc} k_B T$ are defined as follows [3]:

$$C \equiv f_c - \sum_{j=1}^{m} y_{c,j} (\hat{\mu}_{1,j}^o + k_B T \ln x_j) \qquad (2.15)$$

$$E_{sc} k_B T \equiv n_s (f_s - f_c) - n_s \sum_{j=1}^{m} (y_{s,j} - y_{c,j})(\hat{\mu}_{1,j}^o + k_B T \ln x_j) \qquad (2.16)$$

The relation $n_c = k - n_s$ has been used. $C$ is independent of the properties of the endcaps, which are taken into account by $E_{sc}$. The quantity $E_{sc} k_B T / n_s$ represents the mean *excess free energy* per surfactant molecule in the spherical endcaps with respect to a surfactant molecule in the cylindrical part of the micelle. (If we formally set $f_s = f_c$ and $y_{s,j} = y_{c,j}$, then Eq. (2.16) would give $E_{sc} = 0$).

The breakage of a wormlike micelle to two parts leads to the formation of two new endcaps. Consequently, the excess free energy of the two endcaps, $E_{sc} k_B T$, can be identified with the reversible work for breakage of a long wormlike micelle, called also *free energy of scission* [22,23,51,52].

In the special case of single-component micelles, $y_{s,j} = y_{c,j} = 1$ and, then, Eq. (2.16) yields $E_{sc} = n_s (f_s - f_c)/(k_B T)$, which coincides with the definition for $E_{sc}$ in Ref. [1].

Furthermore, the theoretical derivations follow the same way as in Refs. [1,3] and can be seen therein. For this reason, here we give only the final results for micelle size distribution and for the micelle mean mass aggregation number, $n_M$:

$$X_k = \frac{q^k}{K} , \quad q \equiv \frac{X_1}{X_B} , \quad X_B \equiv \exp\left(\frac{C}{k_B T}\right) \qquad (2.17)$$

$$n_{\text{M}} \approx 2[K(X_{\text{S}} - X_{\text{S}}^{\text{o}})]^{1/2}; \quad K \equiv \exp(E_{\text{sc}}) \tag{2.18}$$

where $E_{\text{sc}}$ is given by Eq. (2.16). Here, $X_{\text{S}}$ is the total surfactant molar fraction in the solution,

$$X_{\text{S}} = X_1 + \sum_{k>1} X_k \tag{2.19}$$

and $X_{\text{S}}^{\text{o}}$ is a constant parameter, which can be approximated with the surfactant molar fraction at the CMC [1,35,37].

We recall that Eqs. (2.17) and (2.18) hold for sufficiently large spherocylindrical micelles, for which $n_{\text{c}} \gg n_{\text{s}}$. In the presence of ionic surfactants, $E_{\text{sc}}$ (and $K$) includes contributions from the electrostatic interactions (see Section 2.3). If the salt concentration is much higher than the concentration of ionic surfactant (which is fulfilled in many experiments) then solution's ionic strength is determined by the salt. In such a case, at fixed salt concentration $E_{\text{sc}}$ and $K$ turn out to be independent of $X_{\text{S}}$, and $n_{\text{M}}$ grows linearly with $(X_{\text{S}} - X_{\text{S}}^{\text{o}})^{1/2}$; see Eq. (2.18). However, if the concentrations of salt and ionic surfactant are comparable, then $E_{\text{sc}}$ could depend on $X_{\text{S}}$ and the aforementioned linear dependence could be violated [42,43].

The free energy per molecule in the cylindrical part of the micelle depends on the radius of cylinder, $f_{\text{c}} = f_{\text{c}}(R_{\text{c}})$, and the excess energy of the endcaps (the scission energy) depends on the endcap radius $R_{\text{s}}$ and composition, $y_{\text{s},j}$: $E_{\text{sc}} = E_{\text{sc}}(R_{\text{s}}, y_{\text{s},j})$. The equilibrium value of $R_{\text{c}}$ corresponds to the local minimum of the function $f_{\text{c}}(R_{\text{c}})$. Likewise, the equilibrium values of $R_{\text{s}}$ and $y_{\text{s},j}$ correspond to the local minimum of $E_{\text{sc}}$ with respect to these variables. The latter minimum represents also condition for chemical equilibrium between the endcaps and the cylindrical parts of the micelle; for details, see Ref. [3].

*2.3. Single ionic surfactant and salt*

Here, we further specify the thermodynamic expressions for the special case of single ionic surfactant and salt. For simplicity, we assume that the counterions due to the dissociation of surfactant and salt are the same (e.g. the Na$^+$ ions in the case of sodium dodecyl sulfate and NaCl). In this case, Eqs. (2.12) and (2.16) acquire the following simpler form:

$$k_{\text{c},1}f_{\text{c}} = g_{\text{c},k}^{\text{o}} - k_{\text{c},2}\mu_{\text{a},2}, \quad k_{\text{s},1}f_{\text{s}} = g_{\text{s},k}^{\text{o}} - k_{\text{s},2}\mu_{\text{a},2} \tag{2.20}$$

$$E_{\text{sc}}k_{\text{B}}T = n_{\text{s}}(f_{\text{s}} - f_{\text{c}}) \tag{2.21}$$

Here, the subscripts 1 and 2 denote quantities related to the surfactant ion and counterion, respectively. In view of Eq. (2.20), it is convenient to introduce the auxiliary quantity $f$ defined as follows:

$$k_1 f = g_k^o - k_2 \mu_{a,2} \tag{2.22}$$

When calculated for the micelle cylindrical part (or the endcap), $f$ becomes equal to $f_c$ (or $f_s$); $k_1$ and $k_2$ are the numbers of surfactant molecules and bound counterions in the respective part of the micelle. In analogy with the free energy of adsorption layers from an ionic surfactant [53], the free energy of a micelle from ionic surfactant with bound counterions can be presented as a sum of contributions from the surfactant molecules and from the counterions bound in the Stern layer:

$$g_k^o = k_1 (\mu_{a,1}^o + f_\sigma + f_{conf} + f_{hs} + f_{el,1}) + F_2 \tag{2.23}$$

Here, $\mu_{a,1}^o$ is the standard chemical potential of a surfactant ion in the micellar aggregate; $f_\sigma$ is the interfacial tension component; $f_{conf}$ is the chain conformation component; $f_{hs}$ is the headgroup steric repulsion component; and $f_{el,1}$ is the electrostatic component of the free energy of a surfactant ion in the micelle; see Refs. [1,3]; $F_2$ is the free energy of the counterions in the Stern layer. Expressions for $f_\sigma$, $f_{conf}$ and $f_{hs}$ have been derived in our previous paper, Ref. [1]; these expressions are given in Section 3. In view of the analysis in Ref. [4], $f_{el,1}$ can be expressed in the form:

$$f_{el,1} = z_1 e \psi_s - \pi_{el} \tilde{a} \tag{2.24}$$

where $z_1$ is the valence of surfactant ion; $e$ is the elementary electric charge; $\psi_s$ is the electrostatic potential at the micelle surface; $\pi_{el}$ is the electrostatic component of micelle surface pressure; $\tilde{a} = 1/\Gamma_1$ is the area per surfactant molecule in the surface of charges; $\Gamma_1$ is the number of ionizable surfactant headgroups per unit area of this surface.

An expression for the free energy of the counterions in the Stern layer, $F_2$, can be derived from the canonical ensemble partition function for the distributions of $k_2$ counterions that can occupy $k_1$ adsorption sites on the micellar surface [54]:

$$Q = \frac{k_1!}{k_2!(k_1-k_2)!} [\hat{q}(T)]^{k_2} \exp\left(-\frac{z_2 e \psi_s k_2}{k_B T}\right) \tag{2.25}$$

where $\hat{q}(T)$ is a partition function that accounts for internal degrees of freedom of the bound counterion and $z_2$ is its valence. Using Stirling's formula, one obtains [54]:

$$F_2 = -k_B T \ln Q$$
$$= k_2 \mu_{a,2}^o + k_B T [k_2 \ln k_2 + (k_1 - k_2) \ln(k_1 - k_2) - k_1 \ln k_1] + z_2 e \psi_s k_2 \qquad (2.26)$$

where $\mu_{a,2}^o = -k_B T \ln \hat{q}(T)$ is the standard chemical potential of the bound counterions. Their electrochemical potential is [53-55]:

$$\mu_{a,2} = \left(\frac{\partial F_2}{\partial k_2}\right)_{k_1} = \mu_{a,2}^o + k_B T \ln \frac{k_2}{k_1 - k_2} + z_2 e \psi_s \qquad (2.27)$$

The electrochemical potential of the free counterions in the subsurface layer is

$$\mu_2 = \mu_2^o + k_B T \ln a_{2s} + z_2 e \psi_s \qquad (2.28)$$

where $a_{2s}$ is the activity of the subsurface counterions. Thus, from the condition for equilibrium between bound and free counterions, $\mu_{a,2} = \mu_2$, one obtains the Stern counterion adsorption isotherm [50,56]:

$$\theta \equiv \frac{k_2}{k_1} = \frac{K_{St} a_{2s}}{1 + K_{St} a_{2s}}; \quad K_{St} \equiv \exp\left(\frac{\mu_2^o - \mu_{a,2}^o}{k_B T}\right) \qquad (2.29)$$

which is a Langmuir type isotherm; $K_{St}$ is the Stern constant, and $\theta \equiv k_2/k_1$ is the occupancy of the Stern layer with bound counterions.

Substituting Eqs. (2.23), (2.24), (2.26) and (2.27) into Eq. (2.22), after some algebra we obtain:

$$f = \mu_{a,1}^o + f_\sigma + f_{conf} + f_{hs} + f_{el} \qquad (2.30)$$

where

$$f_{el} \equiv f_{el,1} + k_B T \ln(1 - \theta) = k_B T [\Psi_s + \ln(1 - \theta)] - \pi_{el} \tilde{a} \qquad (2.31)$$

$$\Psi_s \equiv \frac{z_1 e \psi_s}{k_B T} \qquad (2.32)$$

$\Psi_s$ is the dimensionless electric potential on micelle surface; $\Psi_s$ is a positive quantity insofar as $z_1$ and $\psi_s$ have the same sign. In this paper, we consider ionic surfactants and salts that are 1:1 electrolytes; then, $z_1 = \pm 1$ and $z_2 = -z_1$. Note that in Eq. (2.22), the terms with $\mu_{a,2}^o$ from Eqs. (2.26) and (2.27) cancel each other.

In Eq. (2.31), $f_{el}$ is the total electrostatic free energy per surfactant molecule in the micelle, which includes contributions from the free energy of the diffuse EDL and from the counterions in the Stern layer. The term $k_B T \Psi_s = z_1 e \psi_s$ is the electrostatic energy of a surfactant ion incorporated in the micelle. The term $-\pi_{el} \tilde{a}$ expresses the contribution from the diffuse EDL around the micelle; see Section 4.4 and Ref. [4]. The term $k_B T \ln(1-\theta)$ represents a contribution from the configurational free energy of the counterions bound in the Stern layer; this term would be zero at $\theta = 0$ (no counterion binding).

The free energy per molecule in the cylindrical part of the micelle (Fig. 1a) is equal to $f$ in Eq. (2.30) estimated for cylindrical geometry:

$$f_c = \mu_{a,1}^o + f_{c,int}; \quad f_{c,int} \equiv f_{c,\sigma} + f_{c,conf} + f_{c,hs} + f_{c,el} \tag{2.33}$$

As already mentioned, the standard chemical potential $\mu_{a,1}^o$ takes into account contributions from internal degrees of freedom of the surfactant molecule, whereas $f_{c,int}$ accounts for the effect of intermolecular interactions; $\mu_{a,1}^o$ is a constant and does not affect the minimization of $f_c$ with respect to the cylinder radius $R_c$; it is sufficient to minimize $f_{c,int}$ (see Section 5.3).

Analogously, from Eqs. (2.21) and (2.30) one obtains an expression for calculating the scission energy of the spherocylindrical micelle:

$$E_{sc} = (E_{sc})_\sigma + (E_{sc})_{conf} + (E_{sc})_{hs} + (E_{sc})_{el} \tag{2.34}$$

$$(E_{sc})_x \equiv \frac{n_s}{k_B T}(f_{s,x} - f_{c,x}), \quad x = \sigma, conf, hs, el \tag{2.35}$$

As usual, the subscripts "c" and "s" denote the values of the respective quantities calculated for the cylindrical part and the spherical endcaps, respectively. The standard chemical potential, $\mu_{a,1}^o$, is the same for $f_c$ and $f_s$ and does not enter the expression for $E_{sc}$, because it is cancelled in the difference $f_s - f_c$; see Eq. (2.21).

As already mentioned, the equilibrium radius of the cylindrical part, $R_c$, is determined by minimization of the function $f_c(R_c)$. The obtained value of $R_c$ is fixed and, next, the function $E_{sc}(R_s)$ is minimized to determine the equilibrium endcap radius $R_s$. The calculation of the free energy components $f_\sigma$, $f_{conf}$, $f_{hs}$ and $f_{el}$, which enter the expression for $f_c(R_c)$ and $E_{sc}(R_s)$ is described in Section 4. Numerical data for the dependencies $f_c(R_c)$ and $E_{sc}(R_s)$ are presented in Section 5. In view of Eq. (2.18), the theoretical prediction of $E_{sc}$ is equivalent to prediction of the surfactant mean mass aggregation number, $n_M$, for each given surfactant concentration $X_S$.

## 3. Increase of oil/water interfacial tension with the rise of salt concentration

As a rule, rodlike and wormlike micelles grow in solutions of ionic surfactants at high) concentrations of added salt, typically above 0.1 M. At such high ionic strengths, it turns out that the effect of increase of interfacial tension, $\sigma$, with the rise of salt concentration affects significantly micelle growth and must be taken into account when estimating the interfacial tension component of free energy, $f_\sigma$. To quantify this effect, we undertook experimental measurements of the dependence of $\sigma$ on salt concentration. The results are reported here.

It is known that the air/water surface tension increases with the rise of salt concentration in the aqueous phase [57]. However, for oil/water interfaces this effect has been scantily studied. The increment of interfacial tension, $\Delta\sigma_{ow}$, which is due to salt dissolved in water, was investigated by Aveyard and Saleem [58] for dodecane/water interfaces at 20 °C in the presence of various electrolytes. Ikeda et al. [59] measured $\Delta\sigma_{ow}$ for hexane/water interfaces as a function of NaCl concentration, pressure, and temperature.

For the needs of the present study, we carried out measurements of dodecane/water interfacial tension at three different temperatures, viz. 20, 30 and 40 °C, and at NaCl concentrations varying from 0.5 to 1.5 M. The pendant drop method was used with DSA 100R apparatus (Krüss, Germany). Drops from electrolyte solution were formed in the oily phase on a metal capillary of outer diameter 1.83 mm. At each moment of time, the software automatically detects the drop profile and fits it with the Laplace equation of capillarity to determine the interfacial tension.

The used dodecane (> 99.0 %, product of TCI, CAS 112-40-3) was purified by passing through a column filled with silica gel and activated magnesium silicate (Florisil), both of them products of Sigma-Aldrich. Up to three consecutive passes were applied in order to obtain pure oil, for which the dodecane/water interfacial tension does not vary with time because of adsorption of admixtures.

The experimental results for the increment of interfacial tension, $\Delta\sigma_{ow}$, vs. the NaCl concentration, $C_{NaCl}$, are shown in Fig. 2. The dependence of $\Delta\sigma_{ow}$ on both salt concentration and temperature is linear and can be fitted with the equation:

$$\Delta\sigma_{ow} = [0.8923 + 0.0191(T - 273.15)]C_{NaCl} \tag{3.1}$$

In our calculations (Sections 5-7), Eq. (3.1) is applied for hydrocarbon chains with $n_C \geq 11$ carbon atoms.

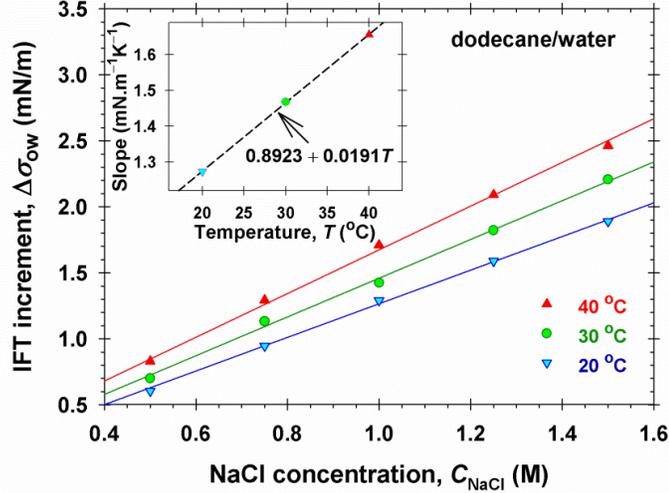

**Fig. 2.** Experimental data for the increment of dodecane/water interfacial tension (IFT), $\Delta\sigma_{ow}$, as a function of the NaCl concentration in the aqueous phase, $C_{NaCl}$, at three temperatures denoted in the figure. The solid lines are linear regressions. The inset shows their slopes plotted vs. temperature, $T$.

**4. Molecular model**

Here, our goal is to calculate the components of micelle free energy, $f_\sigma$, $f_{conf}$, $f_{hs}$ and $f_{el}$, which enter the expressions for the free energy per surfactant molecule, $f_c$, and the micelle scission energy, $E_{sc}$; see Eqs. (2.33) and (2.34). For this goal, we have used information about the molecular properties and some experimental (empirical) dependencies, like Eq. (3.1). In comparison with the case of nonionic surfactants [1], the calculation of $f_\sigma$ and $f_{hs}$ for ionic surfactants takes into account effects related to the specific headgroup shape and the presence of electrolyte in the aqueous phase. The main focus is on the calculation of $f_{el}$.

*4.1. Interfacial tension component of free energy*

The interfacial free energy per surfactant molecule, $f_\sigma$, is due to the contact area of the micelle hydrocarbon core with the outer aqueous phase [1,36]:

$$f_\sigma = \sigma(a - a_0) \tag{4.1}$$

where $a$ is the area per surfactant molecule at the surface of micelle hydrocarbon core; see SI Appendix A (SI = Supplementary Information); $a_0$ is the area, which is excluded (shielded) by the surfactant headgroup; see Fig. 3a, where $a_0 = \pi r_0^2$. It turns out that $f_\sigma$ is the greatest by magnitude component of $f$ (see below), so that $\sigma$ has to be calculated with a high precision, taking into account the effects of (i) surface curvature; (ii) temperature and (iii) added salt.

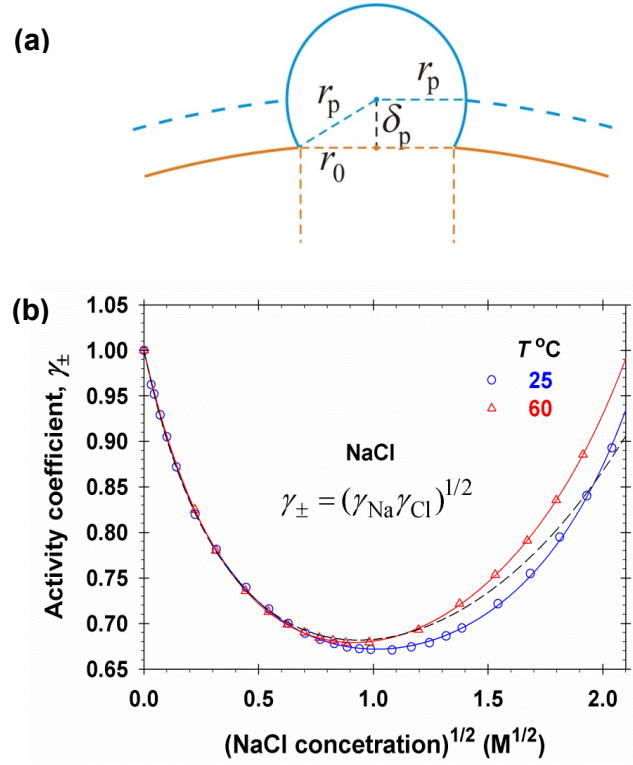

**Fig. 3.** (a) Sketch of the polar headgroup (of radius $r_p$) of an alkyl sulfate molecule; $r_0$ is the radius of the excluded (shielded) area on the surface of micelle hydrocarbon core; $\delta_p$ is the distance between the latter surface and the surface of charges, which is shown by dashed line. (b) Plot of the mean activity coefficient of aqueous NaCl solutions, $\gamma_\pm$, vs. the square root of NaCl concentration, $(C_{NaCl})^{1/2}$, at two temperatures denoted in the figure; the points are experimental data [60]; the solid lines are the best fits with the theoretical model [4]; the dashed line is drawn by using Eq. (4.17).

For this goal, we used the expression [1,61]:

$$\sigma = \frac{\sigma_{ow}}{1+(1/p-1)(\delta_T/R)}; \quad p = \frac{V}{AR} \qquad (4.2)$$

where $p$ is the packing parameter; $V$, $A$ and $R$ are the volume, surface area and curvature radius of the micelle hydrocarbon core: $p = 1/2$ for the cylindrical part of the micelle, whereas $1/3 \leq p \leq 3.8$ for the spherical endcaps; see SI Appendix A; $R = R_c$ for the cylindrical part and $R = R_s$ for the endcaps; see Fig. 1a; $\delta_T$ is the Tolman length [62]; $\sigma_{ow}$ is the interfacial tension of the *flat* oil/water interface, which is estimated from the expression [1]:

$$\sigma_{ow} = 47.12 + 1.479 n_C^{0.5422} - 0.0875(T-293.15) + \Delta\sigma_{ow} \text{ [mN/m]} \qquad (4.3)$$

$n_C$ is the number of carbon atoms in the surfactant alkyl chain; $\Delta\sigma_{ow}$ is given by Eq. (3.1) and accounts for the effect of added salt on $\sigma_{ow}$. In SI Appendix B it is shown that Eq. (4.2) gives values of $\sigma$, which are very close to the exact integral Tolman formula.

There are no general rules how to estimate the Tolman length $\delta_T$. Puvvada and Blankschtein [61] proposed the expression

$$\delta_T = 2.25 \frac{l(n_C)}{l(11)} \text{ [Å]} \tag{4.4}$$

where $l(n_C)$ is the length of alkyl chain with $n_C$ carbon atoms (see Table A1 in SI Appendix A). For $n_C = 12$, Eq. (4.4) yields $\delta_T = 2.43$ Å, which complies very well with the data for the growth of rodlike micelles in solutions of SDS, DDAB and DDAC (see below). It turns out that for sodium unidecyl and tridecyl sulfates ($n_C = 11$ and 13), the experimental data comply very well with the same value, $\delta_T = 2.43$ Å; see Section 6. For TTAB ($n_C = 14$), and CTAB and CTAC ($n_C = 16$), the curvature effect is weaker; in the computations, one could use Eq. (4.4) or set $\delta_T = 2.43$ Å = const. – the results are not sensitive to the value of $\delta_T$; see Section 7.

The excluded (shielded) area, $a_0$, depends on the type and size of the surfactant headgroup [1]. In the case of sulfate headgroup (Sections 5 and 6), agreement theory/experiment has been achieved identifying the excluded area with the cross-sectional area of the surfactant alkyl chain, $a_0 = \alpha(n_C)$; see Table A1 in SI Appendix A. For the studied cationic surfactants (DDAB, DDAC, TTAB, CTAB, and CTAC), the surfactant polar head contains two or three $CH_3$ groups. In this case, $a_0$ is greater than $\alpha(n_C)$; agreement theory/experiment has been achieved by setting $a_0 = a_p$ in Eq. (4.1), where $a_p$ is the maximal cross-sectional area of the polar headgroup; $a_p = \pi r_p^2$ – see Fig. 3a and Section 7.

*4.2. Chain-conformation component of free energy*

Because of the Brownian motion, the mean-square end-to-end distance of a single polymer chain in an ideal solvent is $L_0 = l_{seg} N^{1/2}$, supposedly the chain is anchored with one of its ends [63]; here $l_{seg}$ is the length of a segment and $N$ is the total number of segments in the chain. However, in the confined space of the micelle hydrophobic interior, the surfactant chains are forced to acquire configurations, which are different from that of a free chain in ideal solvent. (Each surfactant hydrocarbon chain in micelle core is surrounded by other hydrocarbon chains; presence of water in the micellar core is not expected [64].) This leads to a chain-conformation contribution to the micelle free energy, $f_{conf}$, which has been quantified using the formula [1]:

$$\frac{f_{conf}}{k_B T} = \frac{3\pi^2 R^2}{16 l_{seg} l} c_{conf}(p); \quad c_{conf}(p) = \frac{4p^2}{1+3p+2p^2} \quad (4.5)$$

where $l$ is the length of the extended surfactant chain and $p$ is the packing parameter; $R = R_c$ for the cylindrical part and $R = R_s$ for the endcaps. The value $l_{seg} = 4.6$ Å was used for the length of a segment from an alkyl chain [65,66]. Eq. (4.5) was derived in Ref. [1] by using the mean-field approach developed by Semenov [67].

As already mentioned, $p = 1/2$ for the cylindrical part of the micelle, whereas $1/3 \le p \le 3/8$ for the micellar endcaps. At that, the conformation coefficient $c_{conf}$ varies in the range $0.200 \le c_{conf}(p) < 0.234$. For the *endcaps*, $p = p(R_s)$ and it is varied when minimizing the scission energy $E_{sc}(R_s)$ to find its equilibrium value; see SI Appendix A and Section 5.3 for details.

*4.3. Headgroup steric repulsion component of free energy*

The repulsion between the surfactant headgroups at the micelle surface, which is due to their finite size, is accounted for by the headgroup steric repulsion component of the free energy per molecule, $f_{hs}$. This component can be described by the excluded-volume term in the free energy of the two-dimensional van der Waals model [1,3,36]:

$$\frac{f_{hs}}{k_B T} = -\ln(1 - \frac{a_p}{\tilde{a}}) \quad (4.6)$$

where $a_p = \pi r_p^2$ is the maximal cross-section of the surfactant headgroup and $\tilde{a}$ is the area per surfactant molecule in the surface, where $a_p = \pi r_p^2$ is located – see the dashed line in Fig. 3.1. Hence, $\tilde{a}$ is related to $a$ in Eq. (4.1) as follows:

$$\tilde{a} = a(1+\delta_p / R_c) \quad \text{(cylindrical part)} \quad (4.7)$$

$$\tilde{a} = a(1+\delta_p / R_s)^2 \quad \text{(endcaps)} \quad (4.8)$$

where $\delta_p$ is the distance between the surface of the maximal cross-section and the surface of the micelle hydrophobic core.

In the case of *alkyl sulfates*, $r_p$ can be identified with the radius of the hydrated sulfate headgroup. Insofar as the latter can be described as a (truncated) sphere (Fig. 3.1), one obtains:

$$\delta_p = (r_p^2 - r_0^2)^{1/2} \tag{4.9}$$

In the case of *cationic surfactants*, such as alkyl dimethyl and trimethyl ammonium salts (e.g. DDAB and TTAB), the headgroup can be approximated as a portion of cylinder, and then $\tilde{a} = a$ and $a_p = a_0$.

*4.4. Electrostatic component of free energy*

*Electrostatic potential.* The electrostatic component of free energy per surfactant molecule in the micelle, $f_{el}$, was calculated from Eq. (2.31). The dimensionless electrostatic potential $\Psi(r)$ in the EDL around the micelle (with $r$ being the radial coordinate) was determined using a cell model (Fig. 1b), which is briefly presented in SI Appendix C and described in details in Ref. [4]. The main advantage of the cell model is that it adequately takes into account the non-uniform character of the micellar solution (with macroions – particles surrounded by EDL). This model predicts the distribution of the ions in the EDL based on the exact mass balance of all species in the micellar solution [4]; see also Refs. [68-70].

The dimensionless surface potential, $\Psi_s$, which enters Eq. (2.31), is estimated at the surface of charges:

$$\Psi_s = \Psi(R + \delta_p) \tag{4.10}$$

where, as usual, $R$ is the radius of the micelle hydrophobic core; $R = R_c$ for the cylindrical part and $R = R_s$ for the endcaps; see Fig. 3.1.

For the cylindrical part of the micelle (Fig. 1a) $\Psi(r)$ was calculated by solving the electrostatic boundary-value problem for an infinite cylinder. Likewise, for the endcaps (truncated spheres) $\Psi(r)$ was calculated for spherical geometry. In other words, the edge effect of the transitional zone between the cylindrical part and the endcaps has been neglected. The obtained excellent agreement between the model and the experimental data (Sections 5-7) confirms that this edge effect is negligible.

*Electrostatic surface pressure*. Another quantity that enters the expression for $f_{el}$, Eq. (2.31), is the electrostatic component of micelle surface pressure, $\pi_{el}$. It has been calculated from the following integral formulas [4]:

$$\pi_{el} = \frac{k_B T}{8\pi\lambda_B} \int_{R_c+\delta_p}^{R_0} \left(\frac{r}{R_c+\delta_p} + \frac{R_c+\delta_p}{r}\right)\left(\frac{\partial \Psi}{\partial r}\right)^2 dr \quad \text{(cylinder)} \tag{4.11}$$

$$\pi_{el} = \frac{k_B T}{12\pi\lambda_B} \int_{R_s+\delta_p}^{R_0} \left[\frac{r^2}{(R_s+\delta_p)^2} + 2\frac{R_s+\delta_p}{r}\right]\left(\frac{\partial \Psi}{\partial r}\right)^2 dr \quad \text{(endcaps)} \tag{4.12}$$

$R_0$ is the radius of the cell that contains the charged micelle and its EDL; see Fig. 1b and SI Appendix C. The above equations take into account also the effect of surface curvature on $\pi_{el}$; for details see Ref. [4]. In Eqs. (4.11) and (4.12), $\lambda_B$ is the Bjerrum length:

$$\lambda_B = \frac{e^2}{4\pi\varepsilon_0\varepsilon k_B T} \tag{4.13}$$

$\lambda_B = 0.71$ nm for water at 25 °C; $\varepsilon_0$ is the dielectric permittivity of vacuum and $\varepsilon$ is the dielectric constant of the aqueous phase.

As known [71,72], $\varepsilon$ depends on temperature, $T$. The magnitude of this dependence is large enough to essentially affect the theoretically predicted scission energy, $E_{sc}$. The dependence $\varepsilon(T)$ has been taken into account by using the empirical formula [71,72]:

$$\varepsilon = 87.74 - 0.40008T + 9.398\times10^{-4}T^2 - 1.41\times10^{-6}T^3 \tag{4.14}$$

where $T$ is to be given in °C.

*Activity coefficients in the EDL.* The expression for $f_{el}$, Eq. (2.31), contains also the occupancy of the Stern layer by bound counterions, $\theta$. This quantity has been calculated from the Stern isotherm, Eq. (2.29), where the activity of the free counterions in the subsurface layer of the micelle is given by the expression:

$$a_{2s} = \gamma_{2,0}c_{2,0}\exp(\Psi_s) \tag{4.15}$$

Here, $\gamma_{2,0}$ and $c_{2,0}$ are, respectively, the activity coefficient and the concentration of counterions at the periphery of the cell around the micelle, at $r = R_0$ where $\psi = 0$ (Fig. 1b). For calculation of the electric potential $\Psi(r)$ we are using the condition for chemical equilibrium, which states that the electrochemical potential is uniform across the EDL (SI Appendix C):

$$\ln(\gamma_j c_j) - (-1)^j \Psi = \ln(\gamma_{j,0} c_{j,0}) \quad (j = 1, 2, 3) \tag{4.16}$$

Here, $j = 1$ stands for the surfactant ions (e.g. dodecyl sulfate anions); $j = 2$ – for the counterions (e.g. $Na^+$), and $j = 3$ – for the coions due to added salt (e.g. $Cl^-$). The concentration of the ions of each kind and the respective activity coefficient depend on the position in the EDL: $c_j = c_j(r)$ and $\gamma_j = \gamma_j(r)$. Eq. (4.15) is a special case of Eq. (4.16) for the surface of the micelle, $r = R + \delta_p$. In the special case of ideal solution ($\gamma_j = \gamma_{j,0} = 1$), Eq. (4.16) reduces to the conventional Boltzmann equation, $c_j = c_{j,0} \exp[(-1)^j \Psi], j = 1, 2, 3$.

In the calculation of $\gamma_j(r)$ we have taken into account (i) the electrostatic interactions between the ions using a generalization of the Debye-Hückel theory for a *nonuniform* electrolyte solution (the EDL) [73], and (ii) the ion-ion steric interactions by using the theoretical expression following from the Boublik-Mansoori-Carnahan-Starling-Leland (BMCSL) model [46-48]; for details – see SI Appendix D and Ref. [4].

The activity coefficients calculated by using this combined approach are in excellent agreement with experimental data for the activity coefficients of bulk electrolyte solutions. As an illustration, Fig. 3b shows a plot of the mean activity coefficient of an *uniform* NaCl solution, $\gamma_\pm = (\gamma_{Na}\gamma_{Cl})^{1/2}$ vs. the square root of salt concentration, $C_{NaCl}$. The experimental points from Ref. [60] and the solid lines predicted by the used model (SI Appendix D) are in excellent agreement. The non-monotonic variation of the activity coefficients is large enough to affect the scission energy $E_{sc}$. The dashed line shows $\gamma_\pm$ predicted by the semiempirical formula originating from the Debye-Hückel theory [60]:

$$\log_{10} \gamma_\pm = -\frac{A\sqrt{I}}{1+B\sqrt{I}} + \hat{b}I \tag{4.17}$$

where $I$ (M) is the ionic strength of solution of 1:1 electrolyte. For NaCl solutions at $T = 25$ °C, the values of the empirical parameters are [60]: $A = 0.5115$ M$^{-1/2}$, $B = 1.316$ M$^{-1/2}$ and $\hat{b} = 0.055$ M$^{-1}$. As seen in Fig. 3b, Eq. (4.17) works well at $C_{NaCl} \leq 0.6$ M, but at higher salt concentrations it predicts values different from the experimental ones. Despite its empirical and approximate character, Eq. (4.17) correctly illustrates the two opposite tendencies that determine the value of activity coefficient and cause its non-monotonic behavior (the minimum). The first negative term in the right-hand side of Eq. (4.17) accounts for the effect of electrostatic ion-ion interactions (Debye screening), which tend to decrease $\gamma_\pm$. The last positive term, $\hat{b}I$, approximately accounts for the effect of steric ion-ion interactions, which becomes significant at the higher electrolyte concentrations and tends to increase $\gamma_\pm$.

In our calculations of $E_{sc}$, we did not use eq. (4.17), which is approximate and refers to a uniform electrolyte solution. Instead, for the diffuse EDL around the micelles we used Eq. (4.16) (instead of the conventional Boltzmann equation) with $\gamma_j = \gamma_j(r)$ being the individual activity coefficients of the ions, which vary across the EDL; see SI Appendix D.

*Effect of the headgroup hydration on the Stern constant.* For sodium alkyl sulfates, the positive charge of the bound sodium ion is located relatively close to the negative charge of the sulfate headgroup. With the rise of temperature, the degree of hydration decreases, so that the two ions come closer and their interaction energy markedly increases, which affects the Stern constant, $K_{St}$, in Eq. (2.29) and influences the value of the calculated scission energy, $E_{sc}$; see Section 5. This effect was taken into account by the following formula:

$$K_{St} = v_2 \exp(\frac{\Delta\mu_2^o}{k_B T} + \frac{\lambda_B}{r_p + r_2}) \tag{4.18}$$

Here, $v_2$ is the volume of the hydrated counterion; the second term in the parentheses is the energy (in $k_B T$ units) of Coulombic interaction between the surfactant headgroup of radius $r_p$ (Fig. 3.1) and the bound counterion of radius $r_2$. The term $\Delta\mu_2^o$ takes into account the contribution from non-Coulombic interactions.

For sodium alkyl sulfates, Eq. (4.18) was used with $v_2 = 0.118$ M$^{-1}$, which is the volume of the hydrated Na$^+$ ion. We used the value $r_2 = 0.95$ Å, which is the radius of the bare Na$^+$ ion. The effect of hydration was incorporated in the value of $r_p$, which was determined as a function of $C_{NaCl}$ and $T$ from the data for $E_{sc}$ of SDS micelles as an adjustable parameter, and afterwards applied to interpret the $E_{sc}$ data for sodium alkyl sulfates of other chainlengths; see Sections 5 and 6.

From data for $K_{St}$ of SDS at $T = 25$ °C in Refs. [53,56], we determined $\Delta\mu_2^o/(k_B T) \approx -0.0753$. This value is much smaller than the Coulombic term $\lambda_B/(r_p + r_2)$, which is close to 2. For this reason, for sodium alkyl sulfates we used Eq. (4.18) with $\Delta\mu_2^o/(k_B T) \approx -0.0753$ neglecting the temperature dependence of this small correction term.

In the case of cationic surfactants, such as alkyl dimethyl and trimethyl ammonium salts (e.g. DDAB and TTAB), the counterion cannot closely approach the headgroup charge because of steric hindrance by the $CH_3$ groups. In such case, the Stern constant, $K_{St}$, can be determined as an adjustable parameter from the fit of data for one member of a homologous surfactant series and to be applied for the other members of this series; see Section 7.

## 5. Wormlike micelles in SDS solutions: theory vs. experiment

### 5.1. Determination of scission energy $E_{sc}$ from available experimental data

Here, our goal is to compare the predictions of the molecular thermodynamic model with available experimental data for the mean mass aggregation number of wormlike micelles, $n_M$, in solutions of sodium dodecyl sulfate (SDS). In Fig. 4a we have plotted data for $n_M$ at 40 °C for various NaBr concentrations obtained by SANS measurements [74]. In accordance with Eq. (2.18), $n_M$ is presented as a function of $(X_S - X_S^o)^{1/2}$, where $X_S$ is the surfactant molar fraction in the aqueous solution and $X_S^o$ is the value of $X_S$ at the critical micellization concentration (CMC). From the slope of the straight lines in Fig. 4a, we determined the micelle growth parameter, $K$, and the micelle scission energy in $k_BT$ units, viz. $E_{sc} = \ln K$. The regression coefficient of the fits in Fig. 4a is $\geq 0.9998$ and the standard error of the determined $E_{sc}$ values in Table 1 is $\pm 0.02$.

**Table 1**. Scission energy $E_{sc}$ ($k_BT$ units) of wormlike SDS micelles determined from the fits of data in Fig. 4a for $T = 40$ °C at various NaBr concentrations.

| NaBr (M) | $E_{sc}$ |
|----------|----------|
| 0.6      | 16.69    |
| 0.7      | 18.40    |
| 0.8      | 19.65    |
| 0.9      | 20.67    |
| 1.0      | 21.80    |

Table 2 presents values of $E_{sc} = \ln K$ obtained by light scattering from SDS solutions with wormlike micelles at various temperatures and NaCl concentrations [38,75]. The second column in Table 2 gives the molar concentration of surfactant in micellar form, $C_S - C_S^o = (X_S - X_S^o)C_W$, where $C_W \approx 55.56$ M is the molar concentration of water.

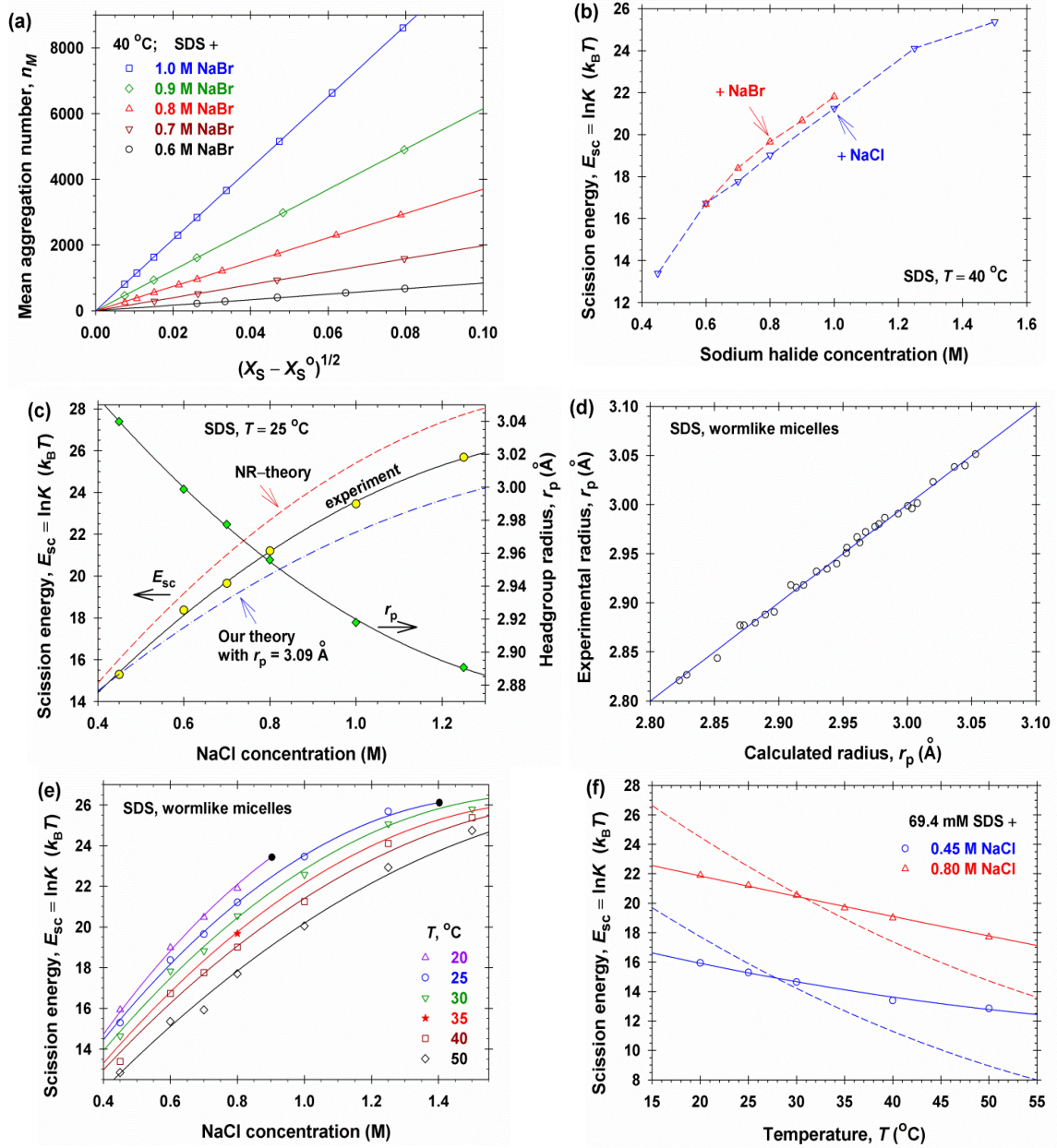

**Fig. 4.** Results for wormlike micelles from SDS. (a) Plot of data from Ref. [74] for the experimental micelle mean mass aggregation number, $n_M$, vs. $(X_S - X_S^\circ)^{1/2}$ in accordance with Eq. (1.1) for micelles of SDS at $T = 40$ °C and at various NaBr concentrations denoted in the figure. (b) Plot of the data for scission energy $E_{sc}$ vs. the concentration of added salt (NaBr or NaCl) from Tables 1 and 2 at $T = 40$ °C. (c) Plot of data from Table 2 at $T = 25$ °C for $E_{sc}$ vs. the NaCl concentration, $C_{NaCl}$; the upper dashed line is the respective theoretical curve from Ref. [36]; the lower dash-dot line is the prediction of our model using the value $r_p = 3.09$ Å determined in the case without added NaCl [53]; the two solid lines, for $r_p$ and $E_{sc}$, are drawn using Eq. (5.1); the points for $r_p$ vs. $C_{NaCl}$ are calculated from the experimental points for $E_{sc}$ using the model from Section 4. (d) Test of Eq. (5.1) against the data at *all* temperatures (Table 2): plot of $r_p$, determined from the experimental $E_{sc}$ at various $T$ and $C_{NaCl}$ vs. $r_p$ calculated from Eq. (5.1); the line is the bisector of first quadrant. (e) Plot of $E_{sc}$ vs. $C_{NaCl}$ for all data in Table 2; the solid lines are predicted by theory using Eq. (5.1); the lines for $T = 20$ and 25 °C terminate with endpoints corresponding to loss of equilibrium between the micelle endcaps and the cylindrical part. (f) Comparison of our theoretical model (the solid lines) with theoretical curves from Ref. [36] (the dashed lines) against experimental data from Table 2.

**Table 2**. Scission energy $E_{sc}$ ($k_B T$ units) of wormlike SDS micelles at various temperatures, surfactant and salt concentrations (light scattering data) [38].

| $T$, °C | $C_S - C_S^o$ (mM) | NaCl (M) | $E_{sc}$ |
|---|---|---|---|
| 20 | 69.4 | 0.45 | 15.93 |
| 20 | 69.4 | 0.60 | 18.99 |
| 20 | 69.4 | 0.70 | 20.48 |
| 20 | 69.4 | 0.80 | 21.90 |
| 25 | 69.4 | 0.45 | 15.29 |
| 25 | 69.4 | 0.60 | 18.37 |
| 25 | 69.4 | 0.70 | 19.65 |
| 25 | 69.4 | 0.80 | 21.21 |
| 25 | 34.7 | 1.00 | 23.45 |
| 25 | 2.32 | 1.25 | 25.68 |
| 30 | 69.4 | 0.45 | 14.64 |
| 30 | 69.4 | 0.60 | 17.84 |
| 30 | 69.4 | 0.70 | 18.83 |
| 30 | 69.4 | 0.80 | 20.55 |
| 30 | 34.7 | 1.00 | 22.58 |
| 30 | 2.32 | 1.25 | 25.07 |
| 30 | 0.694 | 1.50 | 25.79 |
| 35[a] | 38.1 | 0.80 | 19.68 |
| 40 | 69.4 | 0.45 | 13.39 |
| 40 | 69.4 | 0.60 | 16.73 |
| 40 | 69.4 | 0.70 | 17.76 |
| 40 | 69.4 | 0.80 | 19.01 |
| 40 | 34.7 | 1.00 | 21.25 |
| 40 | 2.32 | 1.25 | 24.11 |
| 40 | 0.694 | 1.50 | 25.37 |
| 50 | 69.4 | 0.45 | 12.85 |
| 50 | 69.4 | 0.60 | 15.35 |
| 50 | 69.4 | 0.70 | 15.92 |
| 50 | 69.4 | 0.80 | 17.71 |
| 50 | 34.7 | 1.00 | 20.05 |
| 50 | 2.32 | 1.25 | 22.93 |
| 50 | 0.694 | 1.50 | 24.75 |

[a] Experimental values from Ref. [75].

In Fig. 4b, the values for $E_{sc}$ at 40 °C from Table 1 (SANS, NaBr) and Table 2 (light scattering, NaCl) at 40 °C are compared. In the region of overlapping salt concentrations, the two sets of data for $E_{sc}$ practically coincide. This fact indicates that the effect of salt on micelle growth is dominated by the counterion (Na$^+$), whereas the effect of coion (Cl$^-$ or Br$^-$) is negligible.

*5.2. Effect of the hydrated headgroup radius $r_p$ on $E_{sc}$*

In Fig. 4c, the experimental points (the circles) are data for $E_{sc}$ at 25 °C from Table 2. The upper dashed line is the theoretical prediction from Ref. [36]. The lower dash-dot line, calculated as explained in Section 4, is the prediction of our model, where the value $a_p \equiv \pi r_p^2$ = 30 Å$^2$ ($r_p$ = 3.09 Å) was used for the "equatorial" cross-sectional area of the SDS polar headgroup. This is the value of $a_p$ at not too high NaCl concentrations ($C_{NaCl} \leq 0.115$ M) determined from fits of surface tension isotherms for SDS with the van der Waals model [56,76] and (independently) from electro-conductivity data for SO$_4^{2-}$ ions in aqueous solutions [53].

Calculations based on molecular structure give values of the thermochemical radius of the (bare) sulfate anion in the range $r_{thc}$ = 2.18 – 2.31 Å [77,78]. The greater value $r_p$ = 3.09 Å for aqueous solutions indicates that in water environment the sulfate anion is hydrated. The dehydration of the headgroup of SDS at higher salt concentrations ($C_{NaCl} \geq 0.45$ M in Fig. 4c) could lead to decrease of the effective headgroup radius $r_p$ (salting-out effect) and could cause the difference between the calculated and experimental $E_{sc}$ in Fig. 4c.

To check this hypothesis, we applied the theoretical model from Section 4 in the "opposite direction", viz. to calculate $r_p$ from the experimental $E_{sc}$ values in Fig. 4c. The results for $r_p$ are shown with diamond symbols in the same figure. Note that the whole variation of $r_p$ is in the range 3.04 – 2.89 Å, which is only within 0.15 Å (see the right-hand side vertical axis). However, this small variation of $r_p$ leads to the noticeable difference between the experimental curve and the lower theoretical curve in Fig 4c.

This effect of $r_p$ on $E_{sc}$ can be explained with the dehydration of the sulfate headgroup of SDS with the rise of salt concentration. This effect must be taken into account in order to quantitatively predict the aggregation number $n_M$ of wormlike micelles in aqueous solutions of alkyl sulfates. In addition, dehydration of surfactant headgroups could happen with the rise of temperature, which must be also taken into account. For this goal, we applied the theoretical model from Section 4 to calculate $r_p$ from the values of $E_{sc}$ in Table 2 at various salt concentrations and temperatures. The obtained set of data for $r_p$ at various ionic strengths, $I$, and temperatures, $T$, was fitted with the equation:

$$r_p = a - b(T - 273.15) \ [\text{Å}] \tag{5.1}$$

where $T$ is the absolute temperature; the intercept, $a$, and slope, $b$, depend on the ionic strength $I$:

$$a = 3.292 - 0.5031 I + 0.2414 I^2 \tag{5.2}$$

$$b = 0.003074 - 0.005254 I + 0.004051 I^2 \tag{5.3}$$

In Eqs. (5.2) and (5.3), the dimension of $I$ is moles per liter (M); $I$ is defined as the ionic strength at the border of the cell around each micelle (see Fig. 1b). For wormlike micelles from ionic surfactants, which form in the presence of high salt concentrations (see, e.g., Fig. 4b), $I$ can be approximated with the concentration of added salt.

Note that the considerable variation of $E_{sc}$ with ca. 12 $k_B T$ (see the left-hand side vertical axis in Fig. 4c) is due to the *total* effect of NaCl concentration on the micellar free energy (rather than to the variation of $r_p$ alone). A variation of $\Delta E_{sc} = 12\ k_B T$ leads to variation in micelle aggregation number $n_M$ on the order of $\exp(\Delta E_{sc}/2) \approx 400$ times.

In Fig. 4d, the values of $r_p$, determined from the experimental values of $E_{sc}$ in Table 2 for various NaCl concentrations and temperatures, are plotted vs. $r_p$ calculated from Eqs. (5.1)–(5.3). The plotted data are in excellent agreement with the bisector of the first quadrant, which means that Eqs. (5.1)–(5.3) accurately predict the dependence of sulfate headgroup radius $r_p$ on ionic strength, $I$, and temperature, $T$. In Section 6, Eqs. (5.1)–(5.3), which are obtained by fitting data for SDS, are used to predict the scission energy $E_{sc}$ for wormlike micelles from sodium unidecyl and tridecyl sulfates (SUS and STS).

*5.3. Comparison of theory and experiment and prediction of micellar properties*

Having quantified the effect of electrolyte and temperature on the size of the hydrated sulfate headgroup, we can apply the theory from Section 4 to predict the WLM scission energy, $E_{sc}$, as well as other micellar properties, such as the components of micelle free energy; the equilibrium radii of the cylindrical part and of the spherical endcaps, $R_c$ and $R_s$; the micelle surface electric potential and the occupancy of the Stern layer on micelle surface by bound counterions.

Fig. 4e shows plots of $E_{sc}$ vs. the NaCl concentration at six different temperatures. The symbols are the data from Table 2, whereas the solid lines are predicted by the theory from Section 4, where the headgroup radius $r_p$ is calculated from Eq. (5.1). The used value of the Tolman length is $\delta_T = 2.43$ Å estimated from Eq. (4.4). The presence of excellent agreement between theory and experiment is not surprising insofar as Eq. (5.1) was obtained as a fit based on the same set of data in Table 2 (see Section 5.2). It is interesting that at the higher NaCl concentrations, the theoretical curves at 20 and 25 °C finish with an *endpoint*, denoted by a full black circle. At such higher NaCl concentrations, the theoretical dependence $E_{sc}(R_s)$ has no minimum, which means that the endcaps cannot be in chemical equilibrium with the cylindrical part of the wormlike micelle (see below). In the case of SDS, to the right of the endpoints (Fig. 4e) the experiment shows precipitation (salting out) of surfactant. With ionic wormlike micelles at higher salt concentrations, one could observe also transitions to micelles of another geometry: branched micelles and multiconnected network [8,13,17,18,79] or ribbonlike and disklike micelles [80]. Such transitions explain the experimentally observed peaks in viscosity of micellar solutions vs. the salt concentration – the peaks of the "salt curves" [8,12-17].

It should be noted that the values of $E_{sc}$ in Table 2 (the experimental points in Fig. 4e) have been calculated from the surfactant concentrations, $C_S - C_S^o = (X_S - X_S^o)C_W$, in the same table using Eq. (1.1). Note, however, that the obtained $E_{sc}$ values and the whole theoretical $E_{sc}$ vs. $C_{NaCl}$ curve are *independent* of $C_S - C_S^o$. This is a corollary from the fact that the wormlike micelle is sufficiently long, so that the mean free energy per surfactant molecule in its cylindrical part is independent of micelle length (negligible edge-overlap effects).

To compare the predictions of the model from Section 4 and other previously published models, in Fig. 4f we have plotted $E_{sc}$ vs. $T$ at two different NaCl concentrations, 0.45 and 0.80 M. The symbols are data from Table 2; the solid lines are predicted by the theory from Section 4, whereas the dashed lines are the respective theoretical curves from Ref. [36]. Because $n_M \propto \exp(E_{sc}/2)$, a difference of 10% between the theoretical and experimental $E_{sc}$ values could result in a difference $\geq 300\%$ between the predicted and actual $n_M$ values.

In Figs. 5a, b and c, we compare the different components of the interaction free energy per molecule in the cylindrical part of the micelle, $f_{c,int}$, calculated for three different concentrations of surfactant and salt from Table 2 at 25 °C; see also Eq. (2.33). In these figures, the interfacial tension component $f_{c,\sigma}$ is the greatest one and is the only one which decreases with the rise of micelle radius, $R_c$. The chain conformation component $f_{c,conf}$ is the second greatest and increases with the rise of micelle radius, $R_c$. The electrostatic component $f_{c,el}$ decreases with the rise of NaCl concentration as expected, because the added electrolyte suppresses the electrostatic interactions. At the highest salt concentration, 1.25 mM NaCl (Fig. 5c), $f_{c,el}$ is even negative, which leads to enhancement of micelle growth at this high ionic strength.

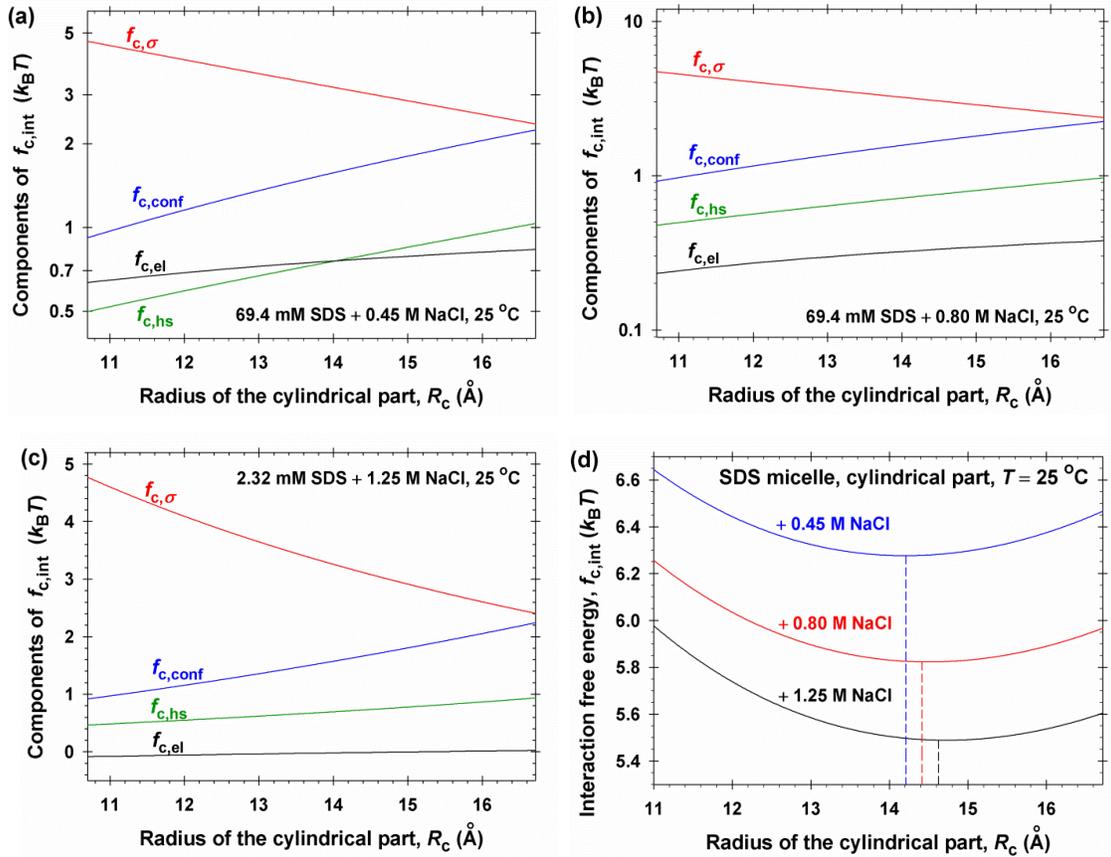

**Fig. 5.** Calculated components of the interaction free energy per molecule in the cylindrical part of SDS micelles, $f_{c,int}$, vs. the radius of cylinder, $R_c$: (a) 0.45 M NaCl; (b) 0.80 M NaCl; (c) 1.25 M NaCl. (d) The total interaction free energy, $f_{c,int} = f_{c,\sigma} + f_{c,conf} + f_{c,hs} + f_{c,el}$, vs. $R_c$; the vertical dashed lines show the positions of the minima of $f_{c,int}$, which determine the respective equilibrium $R_c$ values.

Fig. 5d shows a plot of the total interaction energy, $f_{c,int} = f_{c,\sigma} + f_{c,conf} + f_{c,hc} + f_{c,el}$, vs. $R_c$ for the same three micellar surfactant solutions as in Figs. 5a, b and c. For all these solutions, $f_{c,int}$ has a minimum for $0 < R_c < l$, which determines the *equilibrium* radius of the cylindrical micelle; $l = 16.72$ Å is the length of the extended dodecyl chain of SDS. With the rise of NaCl concentration, $f_{c,int}$ decreases, which favors the formation of wormlike micelles.

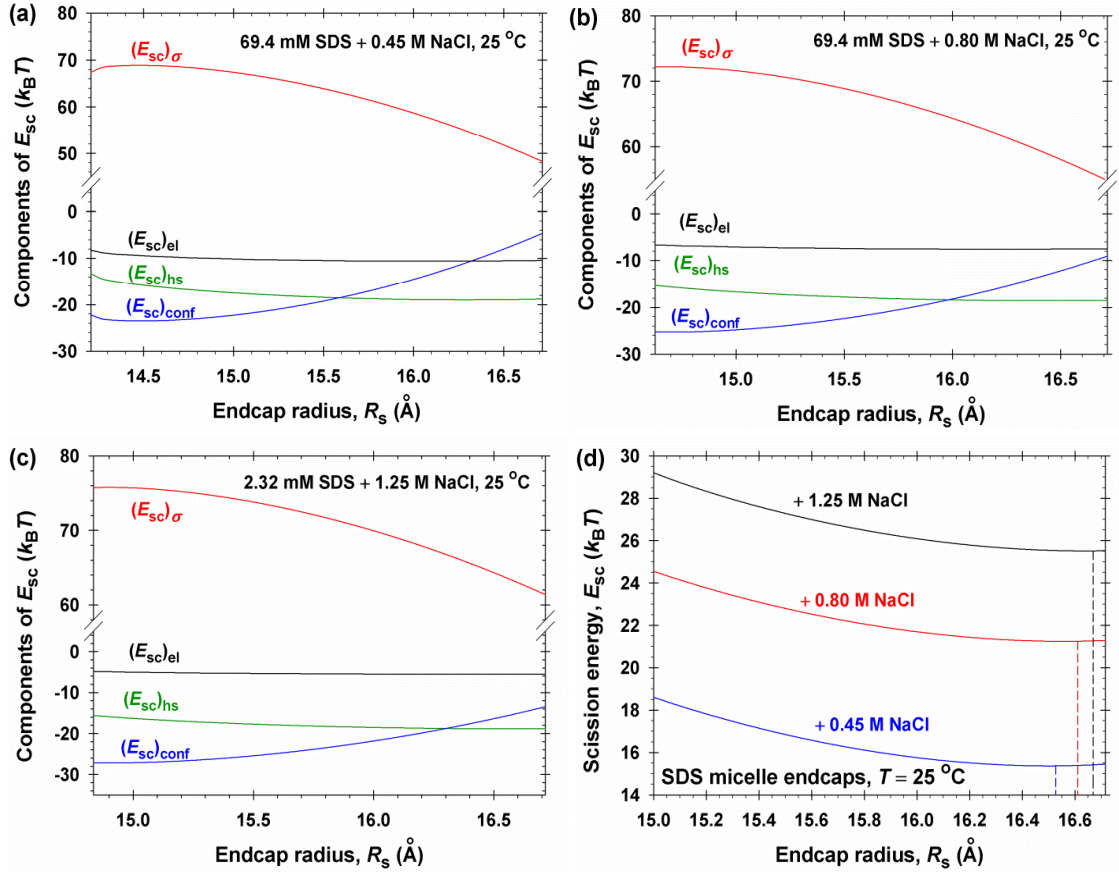

**Fig. 6.** Comparison of the different components of scission energy, $E_{sc}$, plotted as functions of the endcap radius, $R_s$: (a) 0.45 M NaCl; (b) 0.80 M NaCl; (c) 1.25 M NaCl. (d) Plots of $E_{sc} = (E_{sc})_\sigma + (E_{sc})_{conf} + (E_{sc})_{hs} + (E_{sc})_{el}$ vs. $R_s$ for the three NaCl concentrations; the vertical dashed lines show the positions of the minima of $E_{sc}$, which determine the respective equilibrium $R_c$ values.

In Figs. 6a, b and c, we compare the different components of scission energy, $E_{sc}$, which is the excess free energy of the surfactant molecules in the two endcaps of the wormlike micelle with respect to the free energy of the same number of molecules in the cylindrical part. The calculations are performed for the same temperature and salt concentrations as in Figs. 5a, b and c. One sees that the interfacial tension component $(E_{sc})_\sigma$ is the only positive component (that promotes micelle growth) and is considerably greater by magnitude than the other components. In contrast, $(E_{sc})_{conf}$, $(E_{sc})_{hc}$ and $(E_{sc})_{el}$ are negative and oppose the micelle growth. In particular, in most cases the electrostatic component $(E_{sc})_{el}$ is the smallest by magnitude, which could be attributed to the high electrolyte concentration that suppresses the electrostatic interactions.

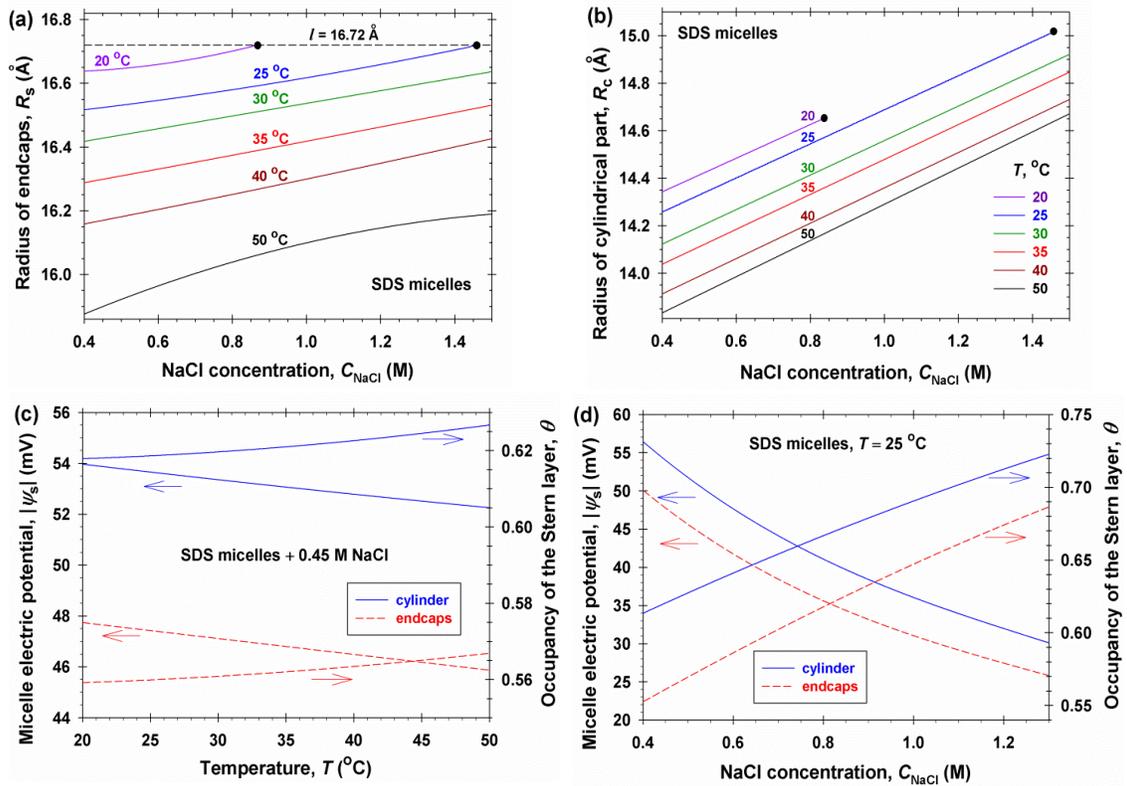

**Fig. 7.** Calculated properties of the equilibrium SDS micelles at various salt concentrations, $C_{NaCl}$, and temperatures, $T$. (a) Plots of the endcap radius, $R_s$, vs. $C_{NaCl}$; $l = 16.72$ Å is the length of the extended dodecyl chain. (b) Plots of the radius of the cylindrical part, $R_c$, vs. $C_{NaCl}$. (c) Plots of micelle surface electrostatic potential, $|\psi_s|$, and of the occupancy of the Stern layer, $\theta$, vs. $T$ for the cylindrical part and the endcaps at 0.45 M added NaCl. (d) Similar plots of $|\psi_s|$ and $\theta$ vs. $C_{NaCl}$ at fixed $T = 25$ °C.

Fig. 6d shows a plot of the total micelle scission energy, $E_{sc} = (E_{sc})_\sigma + (E_{sc})_{conf} + (E_{sc})_{hc} + (E_{sc})_{el}$, vs. the endcap radius $R_s$ for the same three salt concentrations as in Figs. 6a, b and c. With the rise of NaCl concentration, $E_{sc}$ increases, which favors the growth of wormlike micelles. For all these concentrations, $E_{sc}$ has a minimum for $0 < R_s \leq l = 16.72$ Å. The minimum of $E_{sc}$ determines the *equilibrium* value of $R_s$ and corresponds to chemical equilibrium between the endcaps and the cylindrical part of the micelle [1,3]. Note that with the rise of salt concentration, the equilibrium $R_s$ increases and approaches the length of the extended SDS chain, $l = 16.72$ Å. The endpoints of the theoretical curves for 20 and 25 °C in Fig. 4e correspond to minimum, which is located at $R_s = l$. As already mentioned, if there is no minimum of $E_{sc}$ in the physical interval $0 < R_s \leq l$, the endcaps cannot be in chemical equilibrium with the cylindrical part of the wormlike micelle. As already mentioned, at such high salt concentrations one observes precipitation (salting out) of SDS. (For ethoxylated alkyl sulfates, this could be a transition from wormlike to branched micelles [8].)

Figs. 7a and b show the *equilibrium* values of $R_s$ and $R_c$ vs. $C_{NaCl}$ for all calculated curves in Fig. 4e. One sees that both $R_s$ and $R_c$ increase with the rise of salt concentration, but decrease with the rise of temperature. Fig. 7a visualizes the fact that the endpoints in Fig. 4e correspond to endcap radius equal to the extended surfactant chainlength, $R_s = l = 16.72$ Å.

Fig. 7c shows the temperature dependencies of the magnitude of micelle surface potential, $\psi_s$, and the occupancy (with bound $Na^+$ counterions) of the Stern layer at micelle surface, $\theta$, for wormlike SDS micelles in the presence of 0.45 M NaCl. The potential $\psi_s$ is calculated for the surface of charges – the dashed line in Fig. 3a. One sees that both $\psi_s$ and $\theta$ are greater for the cylindrical part as compared to the endcaps. This could be explained with the higher headgroup density at the surface of the cylindrical part. The magnitude of $\psi_s$ slightly increases with the decrease of temperature, $T$. In contract, the counterion binding, $\theta$, is almost independent of $T$: $\theta \approx 59.6$ % for the endcaps and $\theta \approx 65.2$ % for the cylindrical part of the micelle.

Fig. 7d shows the dependencies of $\psi_s$ and $\theta$ on the NaCl concentration at $T = 25$ °C. As expected, the magnitude of surface potential $\psi_s$ decreases, whereas the occupancy of the Stern layer $\theta$ increases with the rise of NaCl. At that, $\theta$ becomes greater than 70% for $C_{NaCl} > 1.2$ M. Again, both $|\psi_s|$ and $\theta$ are greater for the cylindrical part of the micelle as compared to the endcaps.

## 6. Wormlike micelles from SUS and STS: theory vs. experiment

Table 3 shows data for $E_{sc} = \ln K$ determined by Missel et al. [38] by light scattering for wormlike micelles in solutions of *sodium unidecyl* and *tridecyl sulfate* (SUS and STS) at various temperatures and salt concentrations. In Figs. 8a and b, the data for SUS and STS from Table 3 are compared with the theoretical curves for $E_{sc}$ vs. $T$ at the respective NaCl concentrations calculated using the theory from Section 4 with headgroup radius $r_p$ determined from Eq. (5.1) for SDS. The same value of the Tolman length as for SDS, $\delta_T = 2.43$ Å, was used. In other words, the theoretical curves in Figs. 8a and b are drawn without using any adjustable parameters. The obtained excellent agreement between theory and experiment confirms the adequacy of the developed theoretical model.

**Table 3**. Scission energy $E_{sc}$ ($k_B T$ units) of wormlike micelles in solutions of sodium unidecyl sulfate (SUS) and sodium tridecyl sulfate (STS) at various temperatures – light scattering data [38]; the $C_S - C_S^o$ values for SUS or STS are given for each NaCl concentration.

| $T$, °C | $E_{sc}$ |
|---|---|
| 72.9 mM SUS + 0.70 M NaCl | |
| 20 | 15.42 |
| 30 | 14.64 |
| 40 | 13.25 |
| 50 | 12.75 |
| 72.9 mM SUS + 1.00 M NaCl | |
| 20 | 18.57 |
| 25 | 18.11 |
| 30 | 17.48 |
| 35 | 16.89 |
| 40 | 16.21 |
| 50 | 15.27 |
| 36.5 mM SUS + 1.25 M NaCl | |
| 25 | 20.61 |
| 30 | 20.03 |
| 35 | 19.40 |
| 40 | 18.68 |
| 50 | 17.70 |
| 66.1 mM STS + 0.40 M NaCl | |
| 25 | 18.52 |
| 30 | 17.61 |
| 35 | 16.72 |
| 40 | 15.96 |
| 45 | 15.25 |
| 50 | 14.56 |
| 16.5 mM STS + 0.60 M NaCl | |
| 25 | 23.25 |
| 30 | 22.71 |
| 35 | 22.14 |
| 40 | 21.62 |
| 45 | 20.87 |
| 50 | 20.09 |
| 1.65 mM STS + 0.80 M NaCl | |
| 25 | 25.32 |
| 30 | 24.98 |
| 35 | 24.40 |
| 40 | 23.75 |
| 45 | 23.07 |
| 50 | 22.27 |

The fact that agreement between theory and experiment was obtained with $\delta_T$ = 2.43 Å for SUS, SDS and STS indicates that for a homologous series of alkyl sulfates, the Tolman length is independent of the surfactant chainlength, temperature and salt concentration. This is a helpful empirical fact, which can be used for theoretical interpretation and prediction of micelle aggregation number $n_M$. Indeed, it is not easy to theoretically predict $\delta_T$ for ionic surfactant micelles, because $\delta_T$ depends on short range structural effects such as hydration in the headgroup region, discreteness of the surface charge and penetration of the electric field inside the micelle hydrocarbon core.

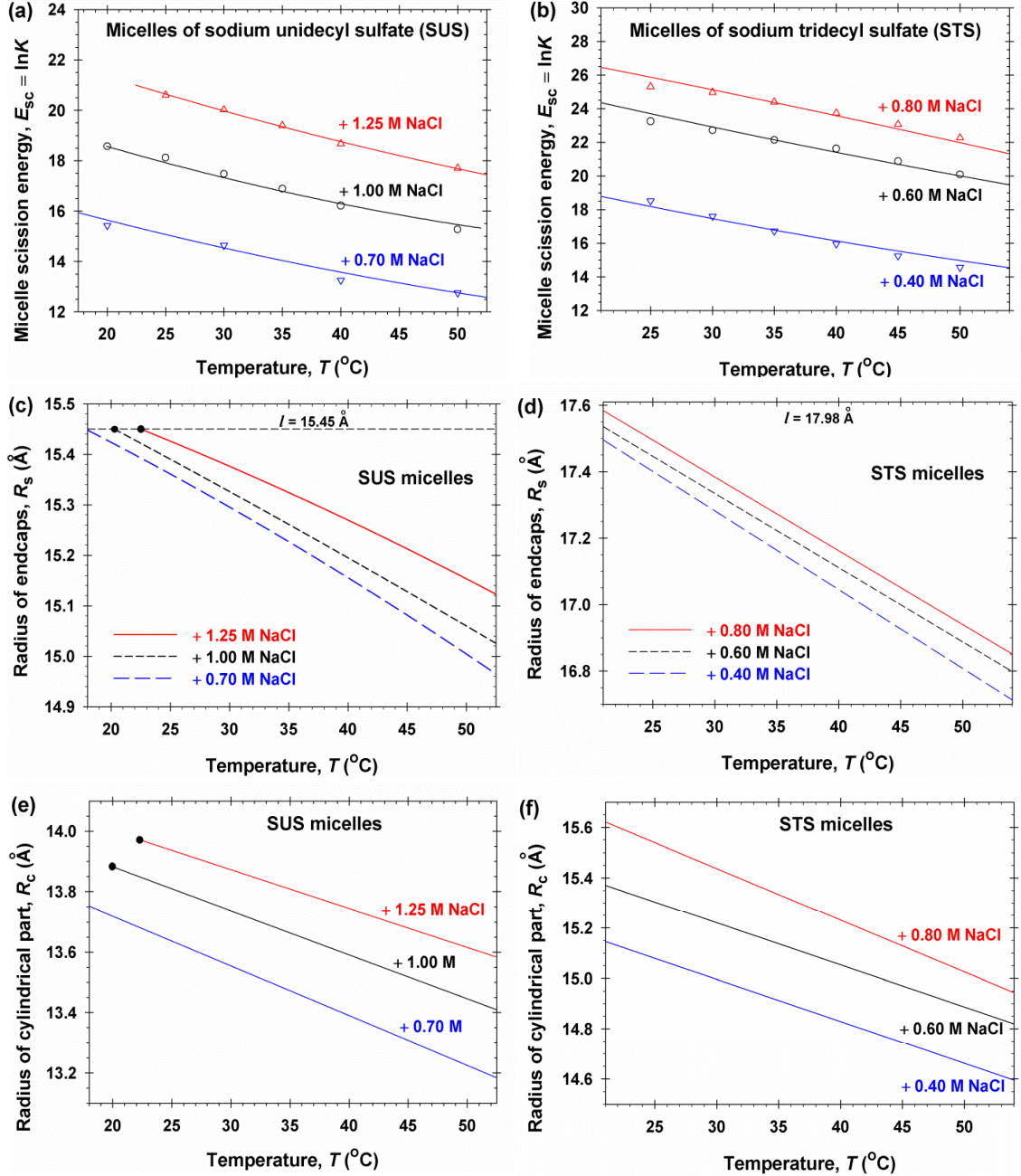

**Fig. 8**. (a) and (b) Plots of $E_{sc}$ vs. $T$ for all data for SUS and STS in Table 3 (the symbols); the solid lines are predicted by the theory in Section 4 without using any adjustable parameters (see the text). (c) and (d) Calculated dependencies of $R_s$ on $T$ for SUS and STS, respectively; $l$ = 15.45 and 17.98 Å are the lengths of the unidecyl and tridecyl chains. (e) and (f) Calculated dependencies of $R_c$ on $T$ for SUS and STS, respectively. The black dots are endpoints corresponding to loss of equilibrium between the micelle endcaps and cylindrical part.

Figs. 8c, d, e and f show the calculated temperature dependencies of the equilibrium radii of micelle endcaps and cylindrical parts, $R_s$ and $R_c$, at various salt concentrations denoted in the figures. The general tendency is $R_s$ and $R_c$ to decrease, i.e. the micelles to become thinner, with the rise of temperature. In contrast, $R_s$ and $R_c$ increase, i.e. the micelles become thicker, with the rise of salt concentration.

For SUS, the curves at the higher salt concentrations (1.00 and 1.25 M NaCl) have endpoints at the lower temperatures; see Figs. 8c and e. The endpoints correspond to $R_s = l = 15.45$ Å, which is the extended length of the unidecyl chain. At temperatures to the left of the endpoints, wormlike micelles should not exist because of the lack of chemical equilibrium between the endcaps and the cylindrical part of the micelles; see the discussions related to Figs. 4e and 6d.

## 7. Wormlike micelles from cationic surfactants: theory vs. experiment

### 7.1. Scission energy determined from experimental data

Here, we compare the predictions of the theoretical model from Section 4 with values of the scission energy of rodlike micelles, $E_{sc}$ (see Table 4), determined from light scattering data for micellar solutions of dodecyldimethylammonium bromide (DDAB) from Ref. [81]; dodecyldimethylammonium chloride (DDAC) from Ref. [82]; tetradecyltrimethylammonium bromide (TTAB) from Ref. [83]; cetyltrimethylammonium bromide (CTAB) from Ref. [84], and cetyltrimethylammonium chloride (CTAC) from Ref. [85]. $E_{sc}$ (in $kT$ units) was calculated from the experimental mean mass aggregation number $n_M$ by means of the equation

$$E_{sc} = \ln\left(\frac{n_M^2 C_W}{4(C_S - C_S^o)}\right) \quad (7.1)$$

see Eq. (1.1). As before, $C_S$ (M) is the total surfactant concentration; $C_S^o$ (M) is the surfactant CMC at the respective salt concentration, and $C_W \approx 55.56$ M is the molar concentration of water. The values of $(C_S - C_S^o)$ are also given in Table 4. In the case of CTAB, the values of $C_S^o$ have been taken from Ref. [86], where the wormlike shape of the micelles has been confirmed by electron microscopy.

**Table 4**. Scission energy $E_{sc}$ ($k_B T$ units) of wormlike micelles in solutions of cationic surfactants at various concentrations of surfactant and salt, with references for the used light scattering data.

| $C_S - C_S^o$ (mM) | Salt (M) | $E_{sc}$ |
|---|---|---|
| DDAB + NaBr at 25 °C [81] | | |
| 61.1 | 0.14 | 16.52 |
| 47.6 | 0.20 | 18.14 |
| 32.3 | 0.30 | 19.85 |
| 17.0 | 0.50 | 22.08 |
| 13.6 | 0.70 | 23.36 |
| 11.9 | 1.00 | 24.46 |
| 11.9 | 1.50 | 25.87 |
| DDAC + NaCl at 25 °C [82] | | |
| 100 | 1.00 | 15.76 |
| 31.2 | 2.00 | 20.49 |
| 8.00 | 3.00 | 24.41 |
| 9.63 | 4.00 | 26.03 |
| TTAB + NaBr at 25 °C [83] | | |
| 47.3 | 0.50 | 18.64 |
| 23.9 | 1.00 | 20.91 |
| 8.83 | 2.00 | 23.79 |
| 5.83 | 3.00 | 25.01 |
| 5.80 | 4.00 | 26.18 |
| CTAB + KBr at 30 °C [84] | | |
| 9.45 | 0.10 | 18.52 |
| 9.59 | 0.20 | 22.51 |
| 9.81 | 0.40 | 24.08 |
| 9.92 | 0.60 | 24.66 |
| CTAC + NaCl at 25 °C [85] | | |
| 62.8 | 1.50 | 16.87 |
| 24.8 | 2.00 | 19.54 |
| 6.41 | 3.00 | 23.75 |
| 6.16 | 4.00 | 26.40 |

*7.2. Results for DDAB and DDAC*

Surface tension isotherms for DDAB and DDAC from Refs. [87,88] are shown in Fig. E1 in SI Appendix E. These surface tension isotherms were fitted with the van der Waals adsorption model augmented with the Stern isotherm of counterion binding as described in Ref. [76] From the fits, we determined the surfactant headgroup area, $a_p$, and the Stern constant, $K_{St}$ – their values are given in Table 5. The considerably greater value of $K_{St}$ for DDAB evidences for a strong binding energy of the $Br^-$ counterions (in comparison with $Cl^-$) to the surfactant headgroups. This was first noticed in Refs. [88,89]. Our calculations (Fig. E2 in SI Appendix E) show that the occupancy of the Stern layer at the CMC levels off at $\theta = 0.75$ for DDAB versus $\theta = 0.61$ for DDAC.

**Table 5**. Molecular parameters for the theoretical curves in Figs. 9a and c for cationic surfactants: cross-sectional area per polar headgroup, $a_p$, and Stern constant of counterion binding, $K_{St}$.

| Surfactant | $T$ (°C) | $a_p$ (Å$^2$) | $K_{St}$ (M$^{-1}$) |
|---|---|---|---|
| DDAB | 25 °C | 37.1 | 1.48 |
| DDAC | 25 °C | 44.6 | 0.493 |
| TTAB | 25 °C | 47.1 | 1.48 |
| CTAB | 30 °C | 47.1 | 1.48 |
| CTAC | 25 °C | 54.6 | 0.493 |

The effective headgroup cross-sectional area $a_p$ turns out to be markedly smaller for DDAB as compared to DDAC (37.1 vs. 44.6 Å$^2$). A possible explanation could be that in the case of DDAB the surfactant dimethylammonium headgroups might have configurations corresponding to closer packing promoted by the stronger binding of the $Br^-$ counterions. Experimentally, this effect is manifested as a greater slope of the experimental surface tension isotherm of DDAB near the CMC (as compared to DDAC); see Fig. E1 in SI Appendix E.

In Fig. 9a, the symbols are the experimental data for DDAB and DDAC from Table 4. The theoretical curves for $E_{sc}$ vs. salt concentrations in Fig. 9a have been calculated by means of the model from Section 4, with the respective values of $a_p$ and $K_{St}$ determined from surface tension fits (see Table 5). In other words, no adjustable parameters have been used to fit the light-scattering data for $E_{sc}$. The obtained excellent agreement between theory and experiment is another strong evidence in favor of the adequacy of the developed molecular thermodynamic model.

The fact that agreement between theory and experiment was obtained by using a constant (independent of ionic strength) value of the headgroup cross-sectional area $a_p$ indicates the lack of salting out (headgroup dehydration) effects in the considered range of salt concentrations at the working temperature of 25 °C. This could be explained with weak hydration of the dimethylammonium headgroup and/or with steric shielding of the hydrating water molecules by the $CH_3$ groups.

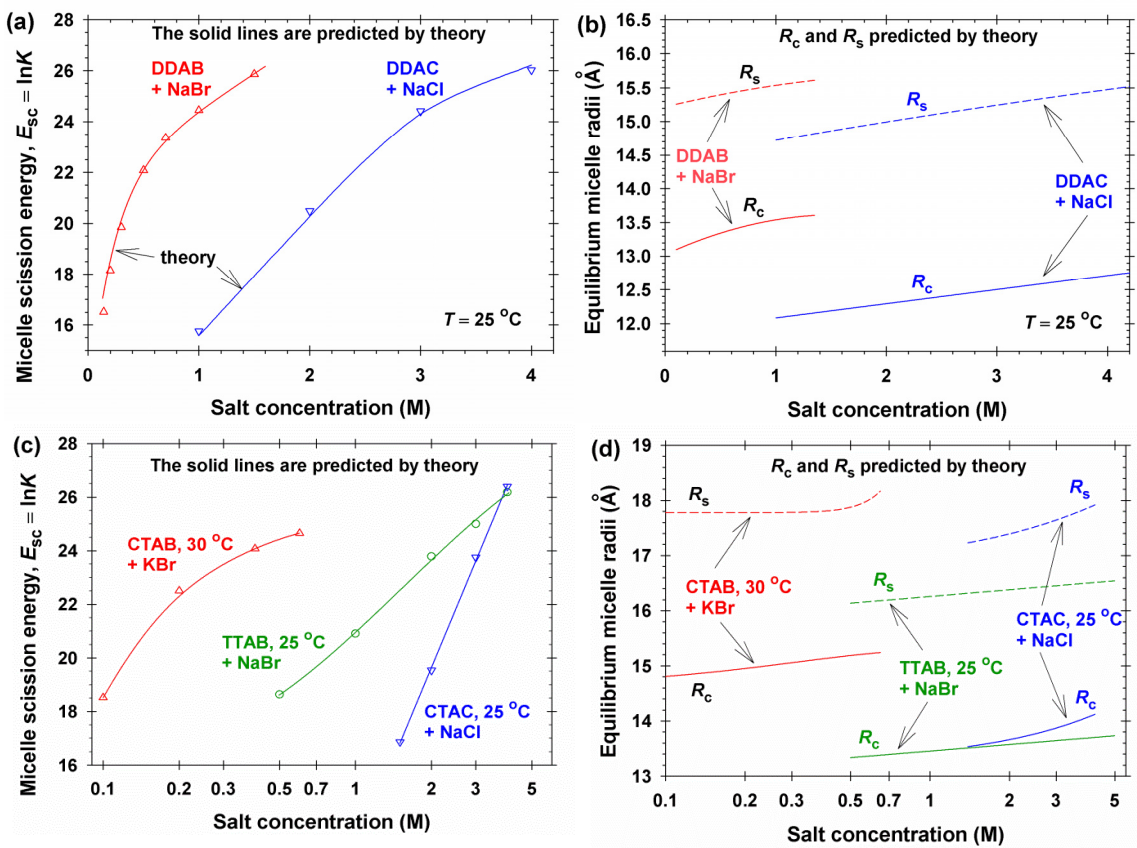

**Fig. 9.** Results for wormlike micelles from cationic surfactants. (a) Plot of the data for scission energy $E_{sc}$ vs. the concentration of added salt (NaBr or NaCl) from Table 4 for DDAB and DDAC; the solid lines are predicted by theory without using any adjustable parameters. (b) Calculated dependencies of the equilibrium cylinder and endcap radii, $R_c$ and $R_s$, on salt concentration for DDAB and DDAC. (c) Plot of the data for $E_{sc}$ vs. the concentration of added salt from Table 4 for CTAB, TTAB and CTAC; the solid lines are predicted by theory using the headgroup cross-sectional area, $a_p$, as a single adjustable parameter; see Table 5. (d) Calculated dependencies of $R_c$ and $R_s$ on salt concentration for CTAB, TTAB and CTAC.

The calculation of $E_{sc}$ by means of the theoretical model includes also calculation of various micellar properties. As an illustration, in Fig. 9b we have plotted the radii of the micelle cylindrical part and of the spherical endcaps, $R_c$ and $R_s$, vs. the salt concentration. One sees that both $R_c$ and $R_s$ are greater for DDAB. This can be explained with the smaller headgroup cross-sectional area, $a_p$, for DDAB, which allows accommodation of more surfactant molecules in the micelle. The values of $R_c$ and $R_s$ in Fig. 9b are calculated in the range of salt concentrations in Table 4. The calculations show that in this range the inequality $0 < R_s < l$ is always satisfied, so that in Fig. 9b there are no endpoints like those in Figs. 4e, 7a and 8c.

*7.3. Results for TTAB, CTAB and CTAC*

In Fig. 9c, the symbols are the experimental data for TTAB, CTAB and CTAC from Table 4. For TTAB and CTAC the data have been obtained at 25 °C [83,85]. The Krafft temperature for CTAB is about 25 °C [90]. The light scattering data for CTAB have been obtained at a higher temperature, 30 °C [84]. The theoretical curves for $E_{sc}$ vs. salt concentration in Fig. 9a have been calculated by means of the model from Section 4.

For TTAB and CTAB, we used the same values of $K_{St}$ as for DDAB, because the surfactant headgroups are similar and the counterions are the same. Likewise, for CTAC we used the same values of $K_{St}$ as for DDAC; see Table 5. The experimental points in Fig. 9c have been fitted with the theory by using the polar headgroup cross-sectional area $a_p$ as a single adjustable parameter. The values obtained from the best fit, $a_p = 47.1$ Å$^2$ for TTAB and CTAB, and $a_p = 54.6$ Å$^2$ for CTAC seem reasonable because of the greater size of the trimethylammonium headgroups as compared to the dimethylammonium ones, and because of the stronger binding of Br$^-$ counterions as compared to the Cl$^-$ ones; see Table 5.

It should be noted that excellent agreement between theory and experiment is obtained for all the three surfactants, TTAB, CTAB and CTAC by using only a single adjustable parameter, $a_p$ (Fig. 9c). At that, the difference between the $a_p$ values for DDAC and DDAB, 44.6 – 37.1 = 7.5 Å$^2$, coincides with the difference between the $a_p$ values for CTAC and CTAB (or TTAB): 54.6 – 47.1 = 7.5 Å$^2$; see Table 5. Probably, this coincidence is not occasional, but reflects the different structural changes in the headgroup region produced by the Cl$^-$ and Br$^-$ ions bound to the micellar surface.

In Fig. 9d, we have plotted the calculated radii of the micelle cylindrical part and of the spherical endcaps, $R_c$ and $R_s$, corresponding to the theoretical curves in Fig. 9c. It is not surprising that $R_c$ and $R_s$ are greater for CTAB as compared to TTAB, because of the longer chain of CTAB. However, it is interesting that $R_c$ is markedly greater for CTAB as compared to CTAC, i.e. the rodlike micelles of CTAB are markedly thicker than those of CTAC despite the identical chainlengths. This fact can be explained with the smaller headgroup area, $a_p$, for CTAB (47.1 vs. 54.6 Å$^2$) which allows incorporation of more surfactant molecules in the equilibrium CTAB micelle as compared to that of CTAC. A similar difference between the $R_c$ values (between the micelle thicknesses) is observed between DDAB and DDAC (Fig. 9b).

## 8. Discussion

*8.1. Comparison of the present study with preceding models*

As already mentioned, to predict the scission energy, $E_{sc}$, and the mean mass aggregation number, $n_M$, of wormlike micelles, one has to calculate the difference between the interaction free energies, $f_s - f_c$, within an accuracy better than $0.01 k_B T$ [4]. To achieve this goal, we strived to avoid (as much as possible) the use of approximations; to keep the number of adjustable parameters to a minimum, and even to work without any adjustable parameters, as in the cases of SUS, STS, DDAB and DDAC. (Otherwise, the comparison of an approximate theory and experiment by adjustment of several unknown parameters acquires empirical character and the model loses its predictive power.) We used known parameters related to the molecular size, or quantitative experimental information, such as the dependences of interfacial tension $\sigma_{ow}$ on the salt concentration and of dielectric constant $\varepsilon$ on temperature, as well as molecular parameters determined from fits of surface tension isotherms (see, e.g., SI Appendix E, Fig. E1).

The correct prediction of electrostatic free energy component, $f_{el} = k_B T[\Psi_s + \ln(1-\theta)] - \pi_{el}\tilde{a}$, demands accurate calculation of the micelle surface electric potential $\Psi_s$, degree of counterion binding, $\theta$, and of the electrostatic component of micelle surface pressure, $\pi_{el}$, the latter being an integral effect over the diffuse electric double layer; see Eqs. (4.11) and (4.12). Because the growth of wormlike micelles from ionic surfactants happens at relatively high salt concentrations (Tables 1–4), the correct calculation of $f_{el}$ demands one to take into account the effect of activity coefficients $\gamma_j$ ($j = 1,2,3$), which have non-monotonic dependence on salinity at high ionic strengths (see, e.g., Fig. 3b). To determine the ionic concentrations, $c_i(r)$, one has to use the electrochemical equilibrium relation, Eq. (4.16) instead of the conventional Boltzmann equation (which corresponds to $\gamma_j = 1$).

Note, that all preceding studies on the prediction of $E_{sc}$ and/or $n_M$ for ionic wormlike micelles, Refs. [36,42-45], are based on an elegant but approximate approach by Ninham et al. [40,41], where the following four approximations have been adopted:

First, in Refs. [40,41] the potential $\psi(r)$ of a given micelle vanishes for $r \to \infty$, where the Debye screening parameter $\kappa$ is determined by the bulk concentrations of salt and dissociated free surfactant molecules. In reality, the extent of the EDL around each micelle is determined by the presence of neighboring micelles. The micellar EDLs are pressed against each other. Here, we have used the more realistic cell model from Ref. [4], where the confinement of micelle EDL by its neighbors is taken into account and accurate mass balances are used for all kinds of ions.

Second, in Refs. [40,41] the Poisson equation is used in combination with the conventional Boltzmann equation, which is equivalent to setting $\gamma_j = 1$ in the electrochemical equilibrium relation, Eq. (4.16). Here, the exact relation, Eq. (4.16), is used along with an expression for the activity coefficient $\gamma_j$, which takes into account the ion-ion interactions from both electrostatic and hard-core origin, and other specific interactions, as well (SI Appendix D, Eqs. (D3)–(D9)). Among the preceding studies, only the model in Ref. [45] includes hard-core interactions between the ions in the free energy expression, but the effect of electrostatic interactions on activity coefficient ($\ln \gamma_i^{(\text{el})}$ in SI Appendix D, Eq. (D4)) is not taken into account.

Third, in Refs. [40,41] the electrostatic free energy of the EDL is calculated using the expression [91]

$$F_{\text{el}} = \int_A \int_0^{\rho_s} \psi_s \, d\rho_s \, dA \tag{8.1}$$

where $A$ is micelle surface and $\rho_s$ is the surface electric charge density. To derive an analytical formula for $F_{\text{el}}$, approximate estimation of integrals, as well as truncated series expansions for small micelle surface curvature have been used [40,41]. Such approximations could essentially affect the calculated values of $E_{\text{sc}}$. In addition, the accurate theory, with $\gamma_j = \gamma_j(r)$ and with a cell model based on mass balances (see SI Appendix C), requires numerical integration with respect to $\rho_s$ in Eq. (8.1). However, this would demand execution of a large number of numerical solutions of the electrostatic boundary-value problem at each step of the numerical integration. Such computational procedure would be too heavy and the accumulation of computational error would be difficult to assess.

In our study, $F_{\text{el}}$ has been calculated based on the expression:

$$\begin{aligned} F_{\text{el}} &= \varepsilon \varepsilon_0 \int_V [\frac{E^2}{2} - \psi \nabla \cdot \mathbf{E} + \int_0^{\psi} (\nabla \cdot \mathbf{E}) \, d\psi] \, dV \\ &= \int_A (\rho_s \psi_s - \pi_{\text{el}}) \, dA \end{aligned} \tag{8.2}$$

where $\mathbf{E} = -\nabla \psi$ is the electric field. The first formula in Eq. (8.2) was derived by Overbeek [92], whereas the last formula in this equation was obtained by us in Ref. [4]. Therein, the equivalence of Eqs. (8.1) and (8.2) is proven in the general case of variable activity coefficients, $\gamma_j = \gamma_j(r)$. For a surface of uniform curvature (sphere, cylinder), Eq. (8.2) reduces to $F_{\text{el}}/A = \rho_s \psi_s - \pi_{\text{el}}$, which allows convenient and accurate calculation of micelle electrostatic free energy with one-time numerical solution of the electrostatic boundary-value problem; see Ref. [4] for details.

Fourth, the effect of counterion binding has not been taken into account in the earlier studies [36,40,41]. Later, Alargova et al. [42,43] demonstrated that the enhancement of micelle growth by divalent and trivalent counterions can be explained only if the counterion binding effect is included in the theory (through the Stern isotherm). This approach was further extended by Srinivasan and Blankschtein [44] to several mono- and multivalent counterions. In Ref. [45], a different approach has been followed, viz. introducing specific characteristics of the solvent-shared surfactant head-counterion pairs [93]. In our study we have used the Stern isotherm of counterion binding, which appears in a natural way in the thermodynamic model, see Eq. (2.29), and leads to a quantitative agreement between the model and the experimental data (Sections 5–7). It should be noted that the Stern isotherm also leads to quantitative agreement between theory and experiment when interpreting data (i) for the effect of salt on the surface tension of ionic surfactant solutions [56,76,94]; (ii) for the effect of salt on the CMC [95,96], and (iii) for the effect of micelles on solutions' conductivity [95].

Moreover, the accurate experimental dependencies $\sigma_{ow}(T,C_{NaCl})$ and $\varepsilon(T)$, see Eqs. (4.3) and (4.14), and the augmented expression for the chain-conformation component of free energy $f_{conf}$, Eq. (4.5), have been first applied to ionic wormlike micelles here.

*8.2. Activity coefficients and dielectric permittivity*

In Ref. [4], the comparison of the theoretical model of the activity coefficients $\gamma_j$ with the experimental data for the mean bulk activity coefficient, $\gamma_\pm = (\gamma_+\gamma_-)^{1/2}$, gave information about the presence of *specific* interactions between the ions in the diffuse EDL. The term "specific interaction" was used for any interaction that is different from the electrostatic and steric (hard-sphere) interactions. Such specific interactions are taken into account by the interaction coefficient $\beta$ in SI Appendix D, Eq. (D10). The comparison of theory and experiment has shown [4] that the specific interactions are relatively small for the ionic pairs $Na^+$–$Cl^-$ and $Na^+$–$Br^-$, whereas the specific interactions become significant ($\beta > 0.1$ M$^{-1}$) for the ionic pairs $K^+$–$Cl^-$ and $K^+$–$Br^-$; see SI Appendix D, Table D1. In particular, the significant difference between the experimental $\gamma_\pm(I)$ dependencies for NaBr and KBr can be explained only if specific interactions in the pair $K^+$–$Br^-$ are taken into account [4].

In the Stern layer on the micelle surface, a considerable part of the water molecules are (partially) immobilized in the hydration shells of the surfactant headgroups and bound counterions, which leads to a lower average dielectric constant, $\varepsilon_s$. This effect has been taken into account for both planar adsorption layers [53] and ionic micelles [97]. It turns out that it can be included in the Stern constant, $K_{St}$. In this way, excellent agreement, between theory and experiment can be achieved for both planar adsorption layers [53,56,76,94,96] and ionic surfactant micelles – in Ref. [95] and here. At that, one can work with the same value of $K_{St}$ for planar interfaces and micelles; see SI Appendix E.

The nonuniform distribution of hydrated ions in the EDL leads to a nonuniform average dielectric permittivity, $\varepsilon(r)$, and even to tensorial dielectric permittivity [98]. In the present study, this effect is taken into account by the variable activity coefficients, $\gamma_j(r)$, of the different kinds of ions, $j = 1,2,3$. In the used model, $\gamma_j(r)$ depends on the local concentrations of *hydrated* ions, which are surrounded by water of bulk dielectric constant $\varepsilon$; see Ref. [4] and SI Appendix D. The achieved good agreement between theory and experiment (see Figs. 4e; 8a and b; 9a and c) indicates that the approach with variable $\gamma_j(r)$ adequately takes into account the non-uniformity of the EDL.

*8.3. Wormlike vs. rodlike micelles*

Another point that needs discussion is whether the experimental light-scattering data analyzed in this paper refer to wormlike or rodlike micelles. (Here, the term "rodlike" is used for straight linear aggregates. Some authors [85] are using the term "semi-flexible rodlike micelles", which is equivalent to wormlike micelles.) In general, the rodlike micelles represent a limiting case of wormlike micelles at small aggregation numbers. The widely used definition is that the linear micelles are rodlike if $L \leq l_p$, where $L$ is the contour length of the micelle and $l_p$ is its persistence length [99]. The light scattering experiments allow one to determine both $L$ and $l_p$, and to compare them. In this way, the authors of Ref. [84] have concluded that the CTAB micelles are rodlike at 0.1 mM KBr, but they are wormlike at 0.2, 0.4 and 0.6 mM KBr (see the data for CTAB in Table 4). For the last row of Table 4, CTAC + 4.0 M NaCl, the authors of Ref. [86] give the following parameter values: $n_M = 11400$; $L = 630$ nm and $l_p = 46.4$ nm; hence, the micelles in this solution are wormlike. The fact that the theory developed in the present article is in excellent agreement with the data for CTAB and CTAC at all studied concentrations (Fig. 9c) means that it predicts correctly $n_M$ (and $L$) for both wormlike and rodlike micelles.

## 9. Conclusions

The growth of spherocylindrical (rodlike, wormlike) micelles is directly related to the excess free energy of the spherical endcaps with respect to the cylindrical part of the micelle, $E_{sc}$, which is known as micelle scission energy. $E_{sc}$ can be determined directly from the measured micelle mean mass aggregation number, $n_M$. In our previous studies [1-3], $E_{sc}$ was calculated for *nonionic* surfactant micelles based on a molecular thermodynamic model and excellent agreement was achieved between theory and experiment. In the present study, our goal was to extend this approach to wormlike micelles from *ionic* surfactants in the presence of added electrolyte (salt). This means (i) to predict the value of $E_{sc}$ knowing the molecular parameters and the input concentrations of ionic surfactant and salt, and (ii) to achieve quantitative agreement between the calculated and measured $E_{sc}$. The present paper is based on the detailed theory of micelle electrostatic free energy developed in Ref. [4], which takes into account the effect of confinement of the EDL of each micelle by the neighboring micelles, as well as the effects of high salt concentrations and finite ionic size through accurate expressions for the activity coefficients, $\gamma_j$. In the previous molecular-thermodynamic studies on the size of ionic surfactant micelles [36,42-45], these effects were either neglected, or they were not incorporated in theory in a self-consistent way; see Section 8.

For this goal, we first extended the thermodynamics of micellization to the case of ionic surfactant micelles taking into account the effect of counterion binding in the Stern layer on the micellar surface (Section 2). The general thermodynamics was developed for a multicomponent system, which may contain several surfactants (both ionic and nonionic) and salts (Sections 2.1 and 2.2). Furthermore, the considerations were focused on a system that consists of single ionic surfactant and salt, which are 1:1 electrolytes (Section 2.3).

The leading component of the ionic micelle scission energy turns out to be the interfacial-tension component, $(E_{sc})_\sigma$, which promotes the micelle growth, whereas the other components of $E_{sc}$ act in the opposite direction; see Fig. 6. For the correct calculation of $(E_{sc})_\sigma$, we determined experimentally the effect of salt concentration on the oil/water interfacial tension $\sigma_{ow}$ at various temperatures (Section 3).

For the correct calculation of the electrostatic component of scission energy, $(E_{sc})_{el}$, we took into account the dependence of dielectric constant of water, $\varepsilon$, on temperature and the variation of ionic activity coefficients, $\gamma_j$, across the EDL (Section 4). As expected, $(E_{sc})_{el}$ is negative (Fig. 6) and tends to diminish the length of the micelles. The increase of $E_{sc}$ (and $n_M$) with the rise of salt concentration (Figs. 4, 8 and 9) is due to the suppression of the effect of $(E_{sc})_{el}$ and to the increase of $(E_{sc})_\sigma$.

To verify the predictions of the theoretical model, we determined and tabulated values of $E_{sc}$ from light scattering data for micellar solutions of anionic and cationic surfactants at various salt concentrations and temperatures; see Tables 1–4. Excellent agreement between the theoretical and experimental values was achieved (Figs. 4e, 7a, and 8c). The model predicts endpoints in the dependencies of $E_{sc}$ on salt concentration or temperature, which indicate loss of balance (lack of chemical equilibrium) between the endcaps and the cylindrical part of the micelle and a transition to self-assemblies of another morphology or onset of crystallization and phase separation [8,13,19].

The obtained explicit analytical expressions give direct quantitative information about the relative importance of various factors on the growth of wormlike micelles in ionic surfactant solutions. The results could be of use for experimentalists, who are planning their experiments and developing new formulations, as well as to theoreticians, who are building up novel models or simulation schemes for complex self-assemblies. The present molecular-thermodynamic approach can be further extended to mixtures of ionic and zwitterionic and/or nonionic surfactants, which may include other amphiphilic molecules, fragrances and preservatives that are contained in the formulations of personal-care and house-hold detergency.

**Acknowledgements**

The authors gratefully acknowledge the support from Unilever R&D, project No. MA-2018-00881N, and from the Operational Programme ''Science and Education for Smart Growth'', Bulgaria, project No. BG05M2OP001-1.002-0023.

**Supplementary Information**

The Supplementary Information to this article contains Appendixes A, B, C, D, E and F.

# Supplementary Information

for the article

## Analytical modeling of micelle growth. 4. Molecular thermodynamics of wormlike micelles from ionic surfactants: theory vs. experiment

Authors: K.D. Danov, P.A. Kralchevsky, R.D. Stanimirova, S.D. Stoyanov, J.L. Cook, and I.P. Stott

Here, the reference numbers are different from those in the main text; the list of cited references is given at the end of the present Supplementary Material.

### Appendix A. Molecular and micellar geometric parameters

For readers' convenience, here we briefly present information, which can be found in our previous paper, Ref. [1], in more detailed form. In Appendix A, all volumes, surface areas, and radii refer to the micelle hydrocarbon core.

*Spherical micelle*: The volume, $V_s$, and the surface area, $A_s$, are related to the radius, $R_s$, as follows:

$$V_s = \frac{4}{3}\pi R_s^3 \; , \; A_s = 4\pi R_s^2 \; , \; p = \frac{V_s}{A_s R_s} = \frac{1}{3} \; , \; N_s = \frac{V_s}{v(n_C)} \tag{A1}$$

where $p$ is the packing parameter and $v(n_C)$ is the volume of an alkyl chain with $n_C$ carbon atoms. The area per surfactant molecule at the surface of micellar core, $a$, is:

$$a = \frac{A_s}{N_s} = \frac{3v(n_C)}{R_s} = \frac{v(n_C)}{R_s p} \tag{A2}$$

*Cylindrical part of a spherocylindrical micelle*: The cylinder volume, $V_c$, its surface area, $A_c$, the packing parameter, $p$, and the number of surfactant molecules contained in the micelle cylindrical part, $n_c$, are related to the radius, $R_c$, and length, $L_c$, of the cylinder as follows:

$$V_c = \pi R_c^2 L_c \; , \; A_c = 2\pi R_c L_c \; , \; p = \frac{V_c}{A_c R_c} = \frac{1}{2} \; , \; n_c = \frac{V_c}{v(n_C)} \tag{A3}$$

The area per surfactant molecule at the surface of micellar core, $a$, is:

$$a = \frac{A_c}{n_c} = \frac{2v(n_C)}{R_c} = \frac{v(n_C)}{R_c p} \tag{A4}$$

*Spherical endcaps of a spherocylindrical micelle*: The endcaps have shapes of truncated spheres with radius $R_s$; the radius of the truncated cross-section is $R_c$ (in general, $R_c \leq R_s$). The expressions for the total volume of the two truncated spheres, $V_{sc}$, the total area of their spherical surfaces, $A_{sc}$, and the total number of surfactant molecules contained in the two endcaps of the micelle, $n_s$, read:

$$V_{sc} = \frac{4\pi}{3} R_s^2 [R_s + (R_s^2 - R_c^2)^{1/2}] + \frac{2\pi}{3} R_c^2 (R_s^2 - R_c^2)^{1/2} \tag{A5}$$

$$A_{sc} = 4\pi R_s [R_s + (R_s^2 - R_c^2)^{1/2}], \quad n_s = \frac{V_{sc}}{v(n_C)} \tag{A6}$$

The packing parameter of the endcaps is defined as follows:

$$p = \frac{V_{sc}}{A_{sc} R_s}, \quad \frac{1}{3} \leq p \leq \frac{3}{8} \tag{A7}$$

The minimum value $p = 1/3$, corresponds to hemispherical endcaps ($R_s = R_c$), whereas the maximum value, $p = 3/8$, corresponds to $R_c / R_s = 3^{1/2}/2 \approx 0.866$.

For the endcaps, the area $a$ per surfactant molecule at the surface of micellar core is:

$$a = \frac{A_{sc}}{n_s} = \frac{A_{sc} v(n_C)}{V_{sc}} = \frac{v(n_C)}{R_s p} \tag{A8}$$

see Eqs. (A5) and (A6). Note that the right-hand sides of Eqs. (A4) and (A8) have analogous form; the differences are in the values of the radius and the packing parameter.

*Alkyl chain parameters*: The volume of alkyl chain is equal to the sum of the volumes of the CH$_3$ and CH$_2$ groups:

$$v(n_C) = v(CH_3) + (n_C - 1)v(CH_2) \tag{A9}$$

where $n_C$ is the number of carbon atoms in the chain. The dependences of $v(CH_3)$ and $v(CH_2)$ on the absolute temperature, $T$, estimated from the temperature dependence of the partial molecular volume of aliphatic hydrocarbons, are:

$$v(CH_3) = 54.3 + 0.124(T - 298.15) \text{ Å}^3 \tag{A10}$$

$$v(CH_2) = 26.9 + 0.0146(T - 298.15) \text{ Å}^3 \tag{A11}$$

Summing up the lengths of the CH$_3$ and CH$_2$ groups, we obtain the length of the extended alkyl chain, $l(n_C)$:

$$l(n_C) = 2.8 + 1.265(n_C - 1) \text{ Å} \tag{A12}$$

The *geometrical constraint* for a spherocylindrical micelle is as follows:

$$R_c \leq R_s \leq l(n_C) \tag{A13}$$

To estimate the cross-sectional area of the alkyl chain, $\alpha(n_C)$, we used the expression:

$$\alpha(n_\text{C}) = \frac{v(n_\text{C})}{l(n_\text{C})} \tag{A14}$$

The values of $l(n_\text{C})$, $\alpha(n_\text{C})$ and $v(n_\text{C})$ at $T = 25\ ^\circ\text{C}$ are listed in Table A1. One sees that $\alpha(n_\text{C})$ varies insignificantly with $n_\text{C}$. In our calculations, Eq. (A14) was used, along with Eqs. (A.9) and (A.12) to calculate $\alpha = \alpha(n_\text{C}, T)$.

**Table A1.** Geometrical parameters of alkyl chains at 25 °C.

| $n_\text{C}$ | $l(n_\text{C})$ Å | $\alpha(n_\text{C})$ Å$^2$ | $v(n_\text{C})$ Å$^3$ |
|---|---|---|---|
| 10 | 14.19 | 20.90 | 296.4 |
| 11 | 15.45 | 20.93 | 323.3 |
| 12 | 16.72 | 20.95 | 350.2 |
| 13 | 17.98 | 20.97 | 377.1 |
| 14 | 19.25 | 20.99 | 404.0 |
| 15 | 20.51 | 21.01 | 430.9 |
| 16 | 21.78 | 21.02 | 457.8 |

## Appendix B. Interfacial tension component of micelle free energy

The interfacial free energy per surfactant molecule, $f_\sigma$, is due to the contact area of the micelle hydrocarbon core with the bulk aqueous phase. The general expression for $f_\sigma$ reads [1,2]:

$$f_\sigma = \sigma(a - a_0) \tag{B1}$$

where: $a_0$ is the surface area excluded by the surfactant headgroup; $a$ is given by Eq. (A4) for the cylindrical part, and by Eq. (A8) for the micelle endcaps.

The effect of curvature on the interfacial tension, $\sigma$, is taken into account by the Tolman formula, which was originally derived for *spherical* surfaces [3]:

$$\frac{\sigma}{\sigma_\text{ow}} = \exp[-\int_0^{\delta_\text{T}/R_\text{s}} \frac{2(3+3x+x^2)}{3+2x(3+3x+x^2)} dx] \tag{B2}$$

where $\delta_\text{T}$ is the Tolman length. The analogous result for *cylindrical* surfaces is [4]:

$$\frac{\sigma}{\sigma_\text{ow}} = \exp[-\int_0^{\delta_\text{T}/R_\text{c}} \frac{2+x}{2+2x+x^2} dx] \tag{B3}$$

Here, $\sigma_{ow}$ is the interfacial tension of the flat phase boundary. The first order series expansions of Eqs. (B2) and (B3) for small $\delta_T/R$ read:

$$\sigma \approx \sigma_{ow}(1-2\frac{\delta_T}{R_s}) \text{ (sphere)}; \quad \sigma \approx \sigma_{ow}(1-\frac{\delta_T}{R_c}) \text{ (cylinder)} \tag{B4}$$

In Fig. B1, the symbols show the numerical predictions of the exact Eqs. (B2) and (B3). The dashed lines correspond to the truncated expansions given by Eq. (B4). One sees that even at $\delta_T/R = 0.1$ the truncated expansions give non-negligible deviations from the exact formula. In our calculations, we used the following more accurate approximated expression [1,5]:

$$\sigma = \sigma_{ow}[1+(\frac{1}{p}-1)\frac{\delta_T}{R}]^{-1} \tag{B5}$$

where $p$ is the packing parameter ($p = 1/3$ for spherical and $p = 1/2$ for cylindrical interfaces). The solid lines in Fig. B1 are calculated using Eq. (B5). The precision of Eq. (B5) is acceptable for estimation of the curvature effect up to $\delta_T/R_s = 0.2$ for spherical surface and up to $\delta_T/R_c = 0.3$ for cylindrical surface.

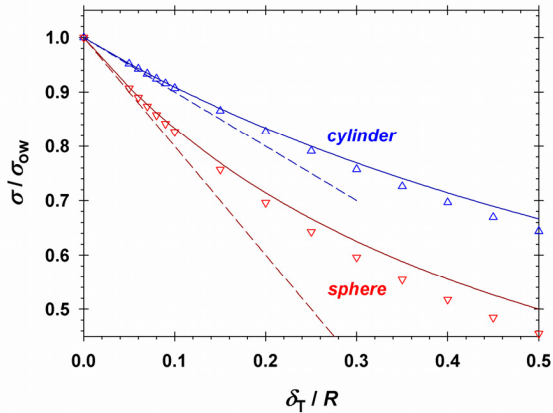

**Fig. B1**. Effect of curvature on the interfacial tension: symbols – Eqs. (B2) and (B3); dashed lines – Eq. (B4); solid lines – Eq. (B5).

The Tolman length can be estimated from the empirical expression [1,5]:

$$\delta_T = 2.25\frac{l(n_C)}{l(11)} \text{ Å} \tag{B6}$$

For example, for $n_C = 12$ one calculates $\delta_T = 2.43$ Å. In our calculations, we used $\delta_T = 2.43$ Å to process experimental data for the surfactants with dodecyl chains, viz. SDS [6], DDAB [7], and DDAC [8]. In the case of SUS [9] and STS [10], the best agreement with the experiment was obtained using the same value, $\delta_T$ = 2.43 Å, as for the dodecyl chains. The effect of electrostatic interactions on the Tolman length is discussed in Ref. [11].

In the case of ionic surfactants of longer alkyl chains, TTAB [12], CTAB [13], and CTAC [14], $\sigma$ is not sensitive to the value of $\delta_T$. For these surfactants, Eq. (B6) has been used to calculate the Tolman length; the obtained values are as follows: $\delta_T$ = 2.80 Å for TTAB, and 3.17 Å for CTAB and CTAC. For these surfactants, if we substitute $\delta_T$ = 2.43 Å (corresponding to dodecyl chains), the obtained changes in the calculated scission energy are less than 2%.

## Appendix C. Calculation of the electrostatic potential

The cell model developed in Ref. [15] is applied. Here, we briefly present the formulation of the problem and the principles of the computational procedure. Details can be found in Ref. [15].

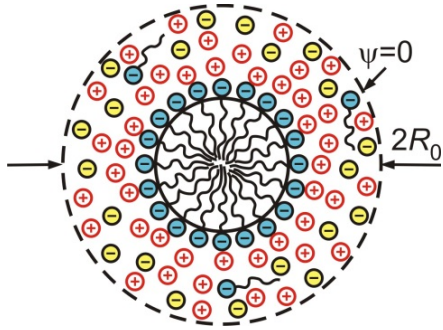

**Fig. C1**. Cross-section of a cylindrical (or spherical) micelle; the used cell model assumes that the electric double layer around the micelle is closed in a cell of outer radius $R_0$, where the electric potential is set zero, $\psi = 0$.

In the cell model, the electrostatic boundary value problem is solved for a cell that contains the micelle and its counterion atmosphere; see Fig. C1. Cylindrical and spherical cells have been used, respectively, for the cylindrical part of the micelle and its endcaps. The outer cell radius, $R_0$, which is different for the cylinder and the endcaps, is determined in the course of the solution of the electrostatic boundary value problem, as explained below.

Within the cell, the Poisson equation can be presented in the form:

$$\frac{1}{r^s}\frac{\partial}{\partial r}(r^s \frac{\partial \psi}{\partial r}) = -\frac{\rho_b}{\varepsilon_0 \varepsilon} \tag{C1}$$

where $\psi$ is the electrostatic potential; $\rho_b$ is the bulk-charge density; $r$ is the radial coordinate; $s = 1$ for cylindrical geometry and $s = 2$ for spherical geometry; $\varepsilon$ is the dielectric constant, and $\varepsilon_0$ is the dielectric permittivity of vacuum. We will consider ionic surfactant and salt, which are 1:1 electrolytes. It is assumed that the counterions due to surfactant and electrolyte are the same (e.g. Na$^+$ ions for SDS and NaCl). In such a case, the bulk charge density, $\rho_b$, and the dimensionless surface potential, $\Psi$, can be presented in the form:

$$\rho_b = q(c_1 - c_2 + c_3) \text{ and } \Psi \equiv \frac{q\psi}{k_B T} > 0 \tag{C2}$$

Here, $q$ is the electric charge of the surfactant ion ($q = +e$ for cationic surfactant and $q = -e$ for anionic one; $e$ is the elementary charge); $c_1$, $c_2$ and $c_3$ are, respectively, the local bulk concentrations of surfactant ions, counterions and coions due to added salt. The area per surfactant molecule on micelle surface, $\tilde{a}$, is inversely proportional to the number of surfactant charges per unit area, $\Gamma_1$, of the *surface of electric charges* of radius $R_{el}$:

$$\Gamma_1 = \frac{1}{\tilde{a}} = \frac{R_c}{aR_{el}}, \quad R_{el} = R_c + \delta_p \quad \text{(cylinder)} \tag{C3}$$

$$\Gamma_1 = \frac{1}{\tilde{a}} = \frac{R_s^2}{aR_{el}^2}, \quad R_{el} = R_s + \delta_p \quad \text{(endcaps)} \tag{C4}$$

Here, $a$ is the area per molecule on the surface of the hydrocarbon core of the micelle. In view of Eq. (C2), we can represent Eq. (C1) in the form:

$$\frac{1}{r^s} \frac{\partial}{\partial r} (r^s \frac{\partial \Psi}{\partial r}) = 4\pi \lambda_B (c_2 - c_1 - c_3) \tag{C5}$$

where $\lambda_B$, is the Bjerrum length:

$$\lambda_B = \frac{e^2}{4\pi \varepsilon_0 \varepsilon k_B T} \tag{C6}$$

Insofar as Eq. (C5) is a second order differential equation, its general solution depends on two integration constants, $A_1$ and $A_2$:

$$\Psi = \Psi(r, A_1, A_2) \tag{C7}$$

The following relations hold at the *outer* border of the cell:

$$\frac{\partial \Psi}{\partial r} = 0 \text{ and } \Psi = 0 \text{ for } r = R_0 \tag{C8a,b}$$

The first relation states that $\Psi$ has a local minimum at the border between two micelles; the second relation, $\Psi(R_0) = 0$, is based on the fact that the electric potential is defined up to an additive constant, which is set zero at the outer cell border. In addition, at the surface of micelle charges (of radius $R_{el}$) the following two relations take place:

$$\Psi_s = \Psi\big|_{r=R_{el}}, \quad \frac{\partial \Psi}{\partial r} = -4\pi\lambda_B(\Gamma_1 - \Gamma_2) \text{ for } r = R_{el} \tag{C9a,b}$$

The first relation is the definition of the dimensionless surface potential $\Psi_s$. The second relation is the dimensionless form of the standard boundary condition relating the normal derivative of potential $\Psi$ with the surface charge density; $\Gamma_1$ is the number of headgroups of the ionic surfactant per unit surface area; $\Gamma_2$ is the number of bound counterions per unit area of micelle surface. $\Gamma_2$ is related to the concentration of counterions by the Stern adsorption isotherm:

$$\frac{\Gamma_2}{\Gamma_1 - \Gamma_2} = K_{St}\gamma_{2,0}c_{2,0}\exp(\Psi_s) \tag{C10}$$

Here, $c_{2,0}$ is the counterion concentration at the outer cell boundary ($r = R_0$); $\gamma_{2,0}$ is the respective activity coefficient; $K_{St}$ is the Stern constant. In fact, $a_{2s} = \gamma_{2,0}c_{2,0}\exp(\Psi_s)$ is the activity of the counterions in the micellar subsurface layer and $\Gamma_2/\Gamma_1 = \theta$ is the occupancy of the Stern layer with bound counterions; see Eq. (2.29) in the main text. The activity coefficient $\gamma_{2,0}$ is calculated as explained in Appendix D (see below).

At equilibrium, the electrochemical potentials are uniform throughout the electric double layer (EDL). In view of Eq. (C8b), this leads to a relation between the ionic concentrations in the EDL, $c_k = c_k(r)$, with their values at the outer cell border, $c_{k,0}$:

$$\ln(\gamma_k c_k) - (-1)^k \Psi = \ln(\gamma_{k,0} c_{k,0}) \quad (k = 1, 2, 3) \tag{C11a,b,c}$$

where $\gamma_k = \gamma_k(c_1,c_2,c_3)$, $k = 1, 2, 3$, are local values of the activity coefficients in the EDL, which are calculated as explained in Appendix D. Correspondingly, $\gamma_{k,0} \equiv \gamma_k(c_{1,0},c_{2,0},c_{3,0})$, $k = 1, 2, 3$. As before, the subscripts 1, 2 and 3 number quantities, which are related, respectively, to surfactant ions, counterions and coions due to added salt; the subscript 0 denotes values of the variables at the outer cell boundary, where $r = R_0$.

In the limiting case of diluted solutions, $\gamma_k = \gamma_{k,0} = 1$, Eqs. (C11a,b,c) are reduced to the Boltzmann equations relating $c_k$ with $\Psi$. However, because the wormlike micelles (WLM) from ionic surfactants grow in relatively concentrated electrolyte solutions, here we always work with $\gamma_k \neq 1$ and $\gamma_{k,0} \neq 1$; see Appendix D.

Finally, to close the system of equations, we have to consider also the mass balances of surfactant and salt. These balances are formulated for the *cylindrical parts* of the micelles, insofar as we consider long micelles, for which the contribution of endcaps to the total mass balance is negligible. Then, the mass balance equations for surfactant ions, counterions and coions due to added salt read [15]:

$$C_1 = (s+1)\frac{\Gamma_1}{R_{el}}(\frac{R_{el}}{R_0})^{(s+1)} + \frac{s+1}{R_0^{s+1}}\int_{R_{el}}^{R_0} c_1 r^s \, dr \tag{C12}$$

$$C_2 = (s+1)\frac{\Gamma_2}{R_{el}}(\frac{R_{el}}{R_0})^{(s+1)} + \frac{s+1}{R_0^{s+1}}\int_{R_{el}}^{R_0} c_2 r^s \, dr \tag{C13}$$

$$C_3 = \frac{s+1}{R_0^{s+1}}\int_{R_{el}}^{R_0} c_3 r^s \, dr \tag{C14}$$

As usual, $s = 1$ for cylinder; $s = 2$ for sphere. The first terms in Eqs. (C12) and (C13) take into account contributions, respectively, from surfactant ions incorporated in the micelles and of counterions bound in the micelle Stern layer. The integral terms in the above three equations take into account contributions from the diffuse part of the EDL, which is located in the domain $R_{el} \leq r \leq R_0$. In Eq. (C14), there is no "adsorption" term because no binding of coions to the (like charged) surfactant headgroups is expected.

In the case of spherocylindrical (wormlike) micelles, these mass balances are to be formulated for the *cylindrical parts* of the micelles ($s = 1$), insofar as we consider long micelles, for which the contribution of the endcaps to the total mass balance is negligible.

Note that Eqs. (C12), (C13) and (C14) are not independent. Indeed, if these equations are substituted in the electroneutrality condition $C_2 - C_1 - C_3 = 0$, and $c_2 - c_1 - c_3$ is substituted from the Poisson equation, Eq. (C5), one obtains Eq. (C9b). Hence, only two among Eqs. (C12), (C13) and (C14) are independent.

For the considered system, the quantity ionic strength, $I$, is defined as follows:

$$I = (c_{1,0} + c_{2,0} + c_{3,0})/2 \tag{C15}$$

## *Computational procedures*

In the procedures (A) and (B) described below, $c_{1,0}$ is an input parameter.

(i) If CMC < $C_1/4$, one can work with $c_{1,0} = 0$. In the case of WLM, it has been verified that this approximation does not affect the first thee significant digits of the calculated $E_{sc}$ value. As usual, CMC is the concentration at which the first (spherical) micelles appear.

(ii) If CMC ≥ $C_1/4$, one has to determine $c_{1,0}$ using the procedures (C) and (D) described below.

(A) *Cylindrical part (s = 1)*. The input parameters are $C_1$, $C_3$, $R_{el}$, $c_{1,0}$, and $\Gamma_1$, the latter determined from Eq. (C3); the Poisson equation, Eq. (C5), is to be solved for $s = 1$ (cylinder). Then, Eqs. (C7), (C8a,b), (C9a,b), (C10), (C11a,b,c), (C12), and (C14) form a system of 11 equations for determining the following 11 unknowns: $\Psi$, $\Psi_s$, $A_1$, $A_2$, $\Gamma_2$, $R_0$, $c_1$, $c_2$, $c_3$, $c_{2,0}$, and $c_{3,0}$. The algorithm for solving this problem can be found in Ref. [15]. The same procedure is applicable also to *spherical* micelles; then, the Poisson equation, Eq. (C5), and Eqs. (C12)–(C14) are to be used for $s = 2$ (sphere).

(B) *Endcaps (s = 2)*. In this case, the concentrations at the outer border of the cell, $c_{2,0}$ and $c_{3,0}$, have been already determined from the solution of the problem for the cylindrical parts of the micelles (see above). The input parameters are $C_1$, $C_3$, $R_{el}$, $\Gamma_1$, $c_{1,0}$, $c_{2,0}$ and $c_{3,0}$. Then, Eqs. (C7), (C8a,b), (C9a,b), (C10), (C11a,b,c) form a system of 9 equations for determining of the following 9 unknowns: $\Psi$, $\Psi_s$, $A_1$, $A_2$, $\Gamma_2$, $R_0$, $c_1$, $c_2$, and $c_3$. The algorithm for solving this problem can be found in Ref. [15].

(C) *Calculations at the CMC of spherical micelles without added salt*. Here, we will use the relation $c_{1,0} = C_W X_1$, where $C_W = 55.56$ M is the molar concentration of water, and $X_1$ is the molar fraction of surfactant. We consider a *spherical* micelle from single ionic surfactant that is 1:1 electrolyte. Then, Eq. (2.7) in the main text (with $m = m_C = 1$ and $x_1 = 1$), which a corollary from the mass action law, reads:

$$\ln X_k = \ln X_1^k - \frac{k}{k_B T}\left(f_{s,int} - \Delta\mu^o_{mic} - k_B T \ln \gamma_1\right) \tag{C16}$$

where $k$ is micelle aggregation number, and

$$f_{s,int} \equiv f_{s,\sigma} + f_{s,conf} + f_{s,hs} + f_{s,el} \tag{C17}$$

$$\Delta\mu_{mic}^o = \mu_{1,1}^o - \mu_{a,1}^o \tag{C18}$$

see also Eqs. (2.3), (2.22) and (2.30) in the main text. Our goal is to determine the difference between the standard chemical potentials, $\Delta\mu_{mic}^o$, at given $X_1$, which is the free-monomer molar fraction at the CMC, which is known from the experimental CMC. For this goal, we first notice that the aggregation number of a spherical micelle

$$k = \frac{(4/3)\pi R_s^3}{v(n_C)} \tag{C19}$$

has a upper limit because the radius of micelle hydrocarbon core, $R_s$, cannot exceed the extended length of the surfactant chain, $l$:

$$k \leq N_{s,max} \equiv \frac{(4/3)\pi l^3}{v(n_C)} \tag{C20}$$

According to Nagarajan and Ruckenstein [2], the CMC can be defined by the relation

$$X_1 = \sum_{k=N_{s,max}/2}^{N_{s,max}} kX_k \tag{C21}$$

($N_{s,max}$ and $N_{s,max}/2$ are to be replaced with the integer parts of the respective numbers). Eq. (C21) states that the amounts of surfactant in micellar and monomeric forms are equal at the CMC; the contribution of micelles with aggregation number $k < N_{s,max}/2$ is neglected; $X_k$ is given by Eq. (C16), along with Eq. (C17). In the latter equation, $f_{s,el}$ is given by the expression:

$$f_{s,el} = k_B T[\Psi_s + \ln(1-\theta)] - \pi_{el}\tilde{a} \tag{C22}$$

see Eq. (2.31) in the main text. All components in Eq. (C17) are calculated as explained in Section 4 of the main article, relative to spherical micelle ($p = 1/3$). In the electrostatic boundary value problem, instead of Eq. (C8a,b), the following boundary condition takes place:

$$\Psi \to 0 \text{ for } r \to \infty \tag{C23}$$

The surfactant and counterion concentrations at $r\to\infty$ are known: $c_{1,0} = c_{2,0} = C_W X_1 =$ CMC. In view of Eq. (C19), for each given aggregation number $k$, we have:

$$R_s = \left(\frac{3v(n_C)k}{4\pi}\right)^{1/3}, \quad \Gamma_1 = \frac{4\pi R_s^2}{k} \frac{1}{(1+\delta_p/R_s)^2} \tag{C24a,b}$$

where $\Gamma_1$ is defined at the surface of charges. Then, Eqs. (C7), (C9a,b), (C10), (C11a,b) and (C23) form a system of 7 equations for determining of the following 7 unknowns: $\Psi$, $\Psi_s$, $A_1$, $A_2$, $\Gamma_2$, $c_1$, and $c_2$. The algorithm for solving this problem can be found in Appendix F.

Furthermore, $f_{s,int}$ in Eq. (C17) is calculated as explained in Section 4 of the main text. Finally, at given $C_W X_1 =$ CMC one solves numerically Eq. (C21), along with Eq. (C16) to determine $\Delta\mu^o_{mic} = \mu^o_{1,1} - \mu^o_{a,1}$. The values of $\Delta\mu^o_{mic}$ determined in this way for the surfactants studied in the present paper are given in Table C1.

**Table C1.** The values of the experimental CMC from the cited references and the standard chemical potential difference $\Delta\mu^o_{mic}$ calculated by us for the studied surfactants.

| Surfactant/[Ref.] | CMC (mM) | $\Delta\mu^o_{mic}/(k_B T)$ |
|---|---|---|
| SUS [31] | 16.0 | 16.82 |
| SDS [31] | 8.08 | 18.26 |
| STS [31] | 4.10 | 19.68 |
| DDAB [Fig. E1] | 11.1 | 16.18 |
| DDAC [Fig. E1] | 14.7 | 16.17 |
| TTAB [32] | 3.86 | 17.85 |
| CTAB [33] | 1.10 | 20.02 |
| CTAC [31] | 1.29 | 20.02 |

One sees that $\Delta\mu^o_{mic}$ for CTAB and CTAC have the same values, which is true also for DDAB and DDAC, for which $\Delta\mu^o_{mic}$ practically coincides. These results are reasonable, because $\Delta\mu^o_{mic}$ is determined by the work of transfer of the surfactant hydrocarbon chain from aqueous in to hydrocarbon environment. The difference between $\Delta\mu^o_{mic}$ for SDS and SUS is 1.44 and between those for STS and SDS – is 1.42, which is quite reasonable because the difference between the respective ionic surfactants is one CH$_2$ group. All these results confirm that the calculation of the interaction part in the expression for the free energy, see Eq. (C17), is accurate, insofar as the Cl$^-$ and Br$^-$ ions have different Stern constants and the surface electrostatic potential and the occupancy of the Stern layer are also different.

(D) *Calculation of $c_{1,0}$ in the case of WLM*. The input parameters are $C_1$, $C_3$, $R_{el}$, $c_{1,0}$, $\Gamma_1$, and $\Delta\mu^o_{mic}$. First a tentative value of $c_{1,0}$ is assumed. Next, we solve the electrostatic boundary value problem: Eqs. (C7), (C8a,b), (C9a,b), (C10), (C11a,b,c), (C12) and (C13) form a system of 11 equations for determining the following 11 unknowns: $\Psi$, $\Psi_s$, $A_1$, $A_2$, $\Gamma_2$, $R_0$, $c_1$, $c_2$, $c_3$, $c_{2,0}$, and $c_{3,0}$. The algorithm for solving this problem can be found in Ref. [15]. Furthermore, interaction free energy per molecule in the *cylindrical* part of the micelle is calculated:

$$f_{c,int} \equiv f_{c,\sigma} + f_{c,conf} + f_{c,hs} + f_{c,el} \tag{C25}$$

see Eq. (2.33) in the main text. Next, from the minimum of $f_{c,int}(R_c)$ we determine the equilibrium values of $f_{c,int}$ and $R_c$. Finally, $c_{1,0}$ is determined by solving numerically the equation expressing the micelle-monomer equilibrium:

$$\Delta\mu^o_{mic} + k_B T \ln(\gamma_{1,0} c_{1,0}) = f_{c,int}(c_{1,0}) \tag{C26}$$

where the standard-chemical-potential difference $\Delta\mu^o_{mic}$ (the same for spherical and cylindrical micelles) has been already determined by using the procedure (C) above; see Table C1.

(E) *Dependence of CMC on the concentration of added salt*. Having determined $\Delta\mu^o_{mic}$ by the procedure (C), one can determine also the dependence of the CMC on the concentration of added salt. In this case, the salt concentration, $c_{3,0}$ is given. First a tentative value of $c_{1,0}$ is assumed and then the counterion concentration at infinity is $c_{2,0} = c_{1,0} + c_{3,0}$. For each $N_{s,max}/2 \leq k \leq N_{s,max}$, we calculate $R_s$ and $\Gamma_1$ from Eqs. (C24a,b). Next, we solve the electrostatic boundary value problem: Eqs. (C7), (C9a,b), (C10), (C11a,b,c), and (C23) form a system of 8 equations for determining the following 8 unknowns: $\Psi$, $\Psi_s$, $A_1$, $A_2$, $\Gamma_2$, $c_1$, $c_2$, and $c_3$. The algorithm for solving this problem can be found in Appendix F. Furthermore, and the interaction free energy per molecule in the spherical micelle, $f_{s,int}$, is calculated as explained in Section 4 of the main text. Finally, $X_1$ (and $c_{1,0} = C_W X_1$) is determined by solving numerically Eq. (C21), along with Eq. (C16) with $\Delta\mu^o_{mic}$ determined by the procedure (C) above; see Table C1.

All procedures described above include determination of all parameters characterizing the respective ionic micelles and, thus, a full quantitative characterization of the micelles is achieved.

**Appendix D. Calculation of activity coefficients**

The chemical potential, $\mu_{1,i}$, of the free ionic species can be presented in the form

$$\mu_{1,i} = \mu_{b,i}^{o} + k_B T \ln(\gamma_i c_i) \quad (i = 1, 2, 3) \tag{D1}$$

The contributions of the inter-ionic interactions to the chemical potential are taken into account by the activity coefficient:

$$k_B T \ln \gamma_i = \mu_i^{(el)} + \mu_i^{(hs)} + \mu_i^{(sp)} \quad (i = 1, 2, 3) \tag{D2}$$

Here, $\mu_i^{(el)}$ takes into account the electrostatic (Debye-Hückel type) interaction between the ions; $\mu_i^{(hs)}$ accounts for the hard-sphere interactions, and finally, $\mu_i^{(sp)}$ expresses the contribution of any other "specific" interactions. Then, the activity coefficient can be presented in the form:

$$\gamma_i = \gamma_i^{(el)} \gamma_i^{(hs)} \gamma_i^{(sp)} \quad (i = 1, 2, 3) \tag{D3}$$

where the three multipliers correspond to the three additives in Eq. (D2).

As before, we use the convention that the subscripts 1, 2 and 3 denote quantities related, respectively, to the surfactant ions, counterions and coions due to the added salt. In fact, Eqs. (D1)–(D3), as well as Eqs. (D4) and (D7) below, are applicable also to an arbitrary number of ionic components, $1 \leq i \leq n$, not necessarily $n = 3$.

To calculate $\gamma_i^{(el)}$, we used the expression [16]:

$$\ln \gamma_i^{(el)} = -\frac{z_i^2 \lambda_B}{b_i} \left[ \frac{\kappa b_i - 2}{2\kappa b_i} + \frac{\ln(1 + \kappa b_i)}{(\kappa b_i)^2} \right]$$

$$-2\pi \frac{z_i^2 \lambda_B^2}{\kappa} \sum_{j=1}^{3} \frac{z_j^2 c_j}{(\kappa b_j)^2} \left[ \frac{2 + \kappa b_j}{1 + \kappa b_j} - \frac{2}{\kappa b_j} \ln(1 + \kappa b_j) \right] \quad (i = 1, 2, 3) \tag{D4}$$

where $b_i$ is the radius of the respective ion (close to its hydrated radius), $z_i$ is its valence, and $\kappa$ is the Debye parameter:

$$\kappa^2 = 4\pi \lambda_B \sum_{i=1}^{3} z_i^2 c_i \tag{D5}$$

Note that in Eqs. (D4) and (D5), $c_i = c_i(r)$ are the local ionic concentrations in the EDL. Consequently, $\kappa = \kappa(r)$ and $\gamma_i^{(el)} = \gamma_i^{(el)}(r)$ also vary across the EDL. In other words, Eq. (D4) generalizes the Debye-Hückel expression to the case of non-uniform solutions. If the ionic radii are equal, $b_1 = b_2 = \ldots = b_n = b$, Eq. (D4) reduces to the simpler formula [16]:

$$\ln \gamma_i^{(\text{el})} = -\frac{z_i^2 \kappa \lambda_B}{2(1+\kappa b)} \tag{D6}$$

For a *uniform* solution, Eq. (D6) coincides with the Debye-Hückel formula for the activity coefficient [17]. Note however, that Eq. (D6) can be used also in a *non-uniform* EDL with $\kappa = \kappa(r)$ and $\gamma_i^{(\text{el})} = \gamma_i^{(\text{el})}(r)$.

To calculate $\gamma_i^{(\text{hs})}$, we used the expression for the activity coefficient of a hard-sphere fluid composed of several components of different radii. This expression, which is derived from the Boublik–Mansoori–Carnahan–Starling–Leland (BMCSL) equation of state, reads [18–20]:

$$\ln \gamma_i^{(\text{hs})} = -\left(1 - 12r_i^2 \frac{\xi_2^2}{\xi_3^2} + 16r_i^3 \frac{\xi_2^3}{\xi_3^3}\right)\ln(1-\xi_3) + \frac{2r_i(3\xi_2 + 6r_i\xi_1 + 4r_i^2\xi_0)}{1-\xi_3}$$

$$+ 12r_i^2 \xi_2 \frac{\xi_2 + 2r_i\xi_1\xi_3}{\xi_3(1-\xi_3)^2} - 8r_i^3 \xi_2^3 \frac{\xi_3^2 - 5\xi_3 + 2}{\xi_3^2(1-\xi_3)^3} \quad (i=1,\ 2,\ 3) \tag{D7}$$

where

$$\xi_m \equiv \frac{\pi}{6}\sum_{i=1}^{3} c_i (2r_i)^m \quad (m = 0,\ 1,\ 2,\ 3) \tag{D8}$$

In Eq. (D8), the index $i$ numbers the components, whereas the index $m$ numbers the powers of the hard-sphere diameter, $(2r_i)$. For the ions in an electrolyte solution, in general, the radii $b_i$ in Eq. (D4) and $r_i$ in Eq. (D7) are different [15]; see Tables D1 and D2 below. In both Eq. (D4) and Eq. (D8), $c_i$ are number (rather than molar) concentrations.

The contribution of other "specific" interactions to the activity coefficient has been quantified by means of the expression [15]

$$\ln \gamma_i^{(\text{sp})} = -2\sum_{j=1}^{3} \beta_{ij} c_j \quad (i=1,\ 2,\ 3) \tag{D9}$$

where the summation is over the different kinds of ions in the solution; $\beta_{ij} = \beta_{ji}$ are interaction parameters. Eq. (D9) can be obtained from the solutions' theory as a *linear approximation* for high (close to 1) molar fraction of the solvent (water), and small molar fractions of the solutes. In this way, Eq. (D9) can be deduced, for example, from the popular Wilson equation; see Eq. (1.200) in Ref. [21]; see also Refs. [22,23].

To avoid using many adjustable parameters, we can further simplify Eq. (D9). Insofar as the like-charged ions repel each other and are separated at greater distances, the significant contribution to $\ln \gamma_i^{(sp)}$ is expected to come from the oppositely charged ions, which can come into close contact. In addition, because in solutions with wormlike micelles the concentration of free surfactant ions is much lower than that of the coions due to salt, a reasonable approximation is $\beta_{12} \approx \beta_{23} \equiv \beta$. Then, Eq. (D9) acquires the following simpler form:

$$\ln \gamma_1^{(sp)} \approx \ln \gamma_3^{(sp)} = -2\beta c_2, \quad \ln \gamma_2^{(sp)} \approx -2\beta(c_1 + c_3) \tag{D10}$$

To determine the values of the radii $b_i$ and $r_i$ for the ions of the used electrolytes, NaCl, NaBr and KBr, in Ref. [15] we fitted literature data for the respective mean activity coefficient $\gamma_\pm = (\gamma_2 \gamma_3)^{1/2}$ (only salt; no surfactant) by using Eqs. (D4), (D7) and (D10). Initially, we varied five adjustable parameters: $b_2$, $b_3$, $r_2$, $r_3$, and $\beta$. The results showed that for the best fit (i) $b_2 \approx b_3 \equiv b$, and (ii) the values of $r_2$ and $r_3$ are very close to the hard-sphere radii of the respective bare ions as given in Ref. [24]. The fact that $b_2 \approx b_3 \equiv b$ probably means that the main contribution to $\gamma_i^{(el)}$ comes from the close contacts in the cationic-anionic pairs, and then $2b$ can be interpreted as the distance between the centers of the ions in such pairs upon contact.

The above result allowed us to fix $r_2$ and $r_3$ equal to the hard-sphere radii of the bare ions in Ref. [24], and to fit the data for $\gamma_\pm$ by using only two adjustable parameters: $b$ and $\beta$. In particular, for $b_2 = b_3 \equiv b$ Eq. (D4) reduces to the simpler Eq. (D6). The results from the best fits of data for $\gamma_\pm$ from Ref. [25] are shown in Table D1 [15].

**Table D1**. Parameters of the model used to calculate the activity coefficients $\gamma_i$ for $i = 2$ and 3: $b$ and $\beta$ are determined from the best fits [15] of literature data [25] for $\gamma_\pm$ for alkali metal halides shown in the first column; $r_+$ and $r_-$ are literature data for the bare ionic radii [24].

| salt | $T$ (°C) | $b$ (Å) | $r_+$ (Å) | $r_-$ (Å) | $\beta$ (M$^{-1}$) |
|---|---|---|---|---|---|
| NaCl | 60° | 4.47 | 1.009 | 1.822 | 0.00466 |
| NaCl | 25° | 3.95 | 1.009 | 1.822 | 0.00966 |
| NaBr | 25° | 4.02 | 1.009 | 1.983 | $\approx 0$ |
| KCl | 25° | 3.90 | 1.320 | 1.822 | 0.0635 |
| KBr | 25° | 4.28 | 1.320 | 1.983 | 0.0789 |

Here, by definition $r_2$ and $r_3$ are the radii of the counterions and coions in a micellar solution, whereas in Table D1 $r_+$ and $r_-$ are radii of cations and anions. Therefore, in the case of *anionic* surfactant, e.g. SDS + NaCl, we have $r_2 = r_+$ and $r_3 = r_-$. Conversely, in the case of *cationic* surfactant, e.g. CTAC + NaCl, we have $r_2 = r_-$ and $r_3 = r_+$.

As already mentioned, to calculate the activity coefficient, $\gamma_1$, of the free surfactant ions, which appear with a low concentration in the EDL (much lower than that of salt), a reasonable approximation is that they can be treated as the coions due to salt (see Table D1) with the only difference that the effective radius, $r_1$, of the surfactant ions is greater. In Table D2, we have summarized the parameter values used to calculate $\gamma_i$ for all studied systems of surfactant + salt, where $r_1$ has been estimated from molecular size considerations.

**Table D2**. Parameters of the model used to calculate the activity coefficients $\gamma_i$ for $i$ = 1, 2, 3 and for all studied surfactant + salt systems.

| System | $T$ (°C) | $b$ (Å) | $r_1$ (Å) | $r_2$ (Å) | $r_3$ (Å) | $\beta$ (M$^{-1}$) |
|---|---|---|---|---|---|---|
| SUS + NaCl | 25° | 3.95 | 4.55 | 1.009 | 1.822 | 0.00966 |
| SUS + NaCl | 60° | 4.47 | 4.55 | 1.009 | 1.822 | 0.00466 |
| SDS + NaCl | 25° | 3.95 | 4.65 | 1.009 | 1.822 | 0.00966 |
| SDS + NaCl | 60° | 4.47 | 4.65 | 1.009 | 1.822 | 0.00466 |
| STS + NaCl | 25° | 3.95 | 4.75 | 1.009 | 1.822 | 0.00966 |
| STS + NaCl | 60° | 4.47 | 4.75 | 1.009 | 1.822 | 0.00466 |
| DDAB + NaBr | 25° | 4.02 | 5.15 | 1.983 | 1.009 | $\approx 0$ |
| DDAC + NaCl | 25° | 3.95 | 5.15 | 1.822 | 1.009 | 0.00966 |
| TTAB + NaBr | 25° | 4.02 | 5.30 | 1.983 | 1.009 | $\approx 0$ |
| CTAB + KBr | 25° | 4.28 | 5.45 | 1.983 | 1.320 | 0.0789 |
| CTAC + NaCl | 25° | 3.95 | 5.45 | 1.822 | 1.009 | 0.00966 |

The values of $\gamma_i$ for SUS, SDS and STS at temperatures 25° ≤ $T$ ≤ 60° have been calculated by linear interpolation between $\gamma_i(25°)$ and $\gamma_i(60°)$:

$$\gamma_i = \gamma_i(25°)\frac{60-T}{35} + \gamma_i(60°)\frac{T-25}{35}, \quad i=1,2,3. \tag{D11}$$

Indeed, as seen in Fig. 3b in the main text, the temperature dependence of the activity coefficient is not so considerable. For this reason, for the system CTAB + KBr at 30 °C we used the parameter values for CTAB + KBr at 25 °C given in Table D2.

**Appendix E. Parameters for DDAB and DDAC obtained from surface tension fits**

For the cationic surfactants DDAB and DDAC, we processed surface tension isotherms (air/water interface) from Refs. [26,27] to obtain the respective values of the headgroup area, $a_p$, and the Stern constant, $K_{St}$. A combined van der Waals type adsorption model for the surfactant molecules and the Stern adsorption isotherm for the bound counterions was used [28,29]. The parameters for the surfactant hydrophobic tails were fixed to be the same as for SDS: energy of surfactant adsorption $E_1 = 12.53\ k_BT$ and dimensionless interaction parameter $\hat{\beta} = 2.73$; see Table 4 in Ref. [29]. The two adjustable parameters, $K_{St}$ and $a_p$, were varied to obtain the best theoretical fits shown in Fig. E1. The calculated parameters are given in Table E1 (see also Table 5 in the main text).

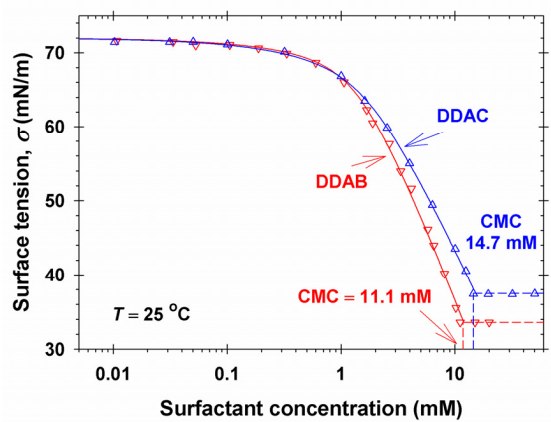

**Fig. E1**. Surface tension isotherms of DDAB and DDAC measured at 25 °C. The symbols correspond to experimental data [26,27]. The solid lines show the best fit according to the theoretical model from Refs. [28,29].

**Table E1**. Molecular parameters for DDAB and DDAC determined from the fits in Fig. E1: cross-sectional area per polar headgroup, $a_p$, and Stern constant of counterion binding, $K_{St}$.

| Surfactant | $T$ (°C) | $a_p$ (Å$^2$) | $K_{St}$ (M$^{-1}$) |
|---|---|---|---|
| DDAB | 25 °C | 37.1 | 1.48 |
| DDAC | 25 °C | 44.6 | 0.493 |

The slope of the surface tension isotherm near the CMC is greater for DDAB, as compared to DDAC (Fig. E1), which leads to smaller $a_p$ and greater $K_{St}$ for DDAB (Table E1). This difference indicates higher binding energy of the Br$^-$ ions (as compared to Cl$^-$) to the headgroups of these cationic surfactants [27,30]. This seems to lead to changes in the conformation and packing of the surfactant molecules in the dense adsorption layers.

With the parameter values determined from the best fit (Table E1), we calculated several properties of the surfactant adsorption layers as predicted by the used theoretical model [28,29] – see Fig. E2. As expected, the surfactant adsorption $\Gamma_1$ is greater for DDAB and the difference with DDAC increases when approaching the CMC. The occupancy of the Stern layer, $\theta = \Gamma_2/\Gamma_1$, is markedly greater for DDAB. Correspondingly, the surface potential magnitude, $|\psi_s|$, is smaller for DDAB because of a partial neutralization of the surface charge by bound counterions. The non-monotonic behavior of $|\psi_s|$ is typical for *ionic* surfactants: At lower concentrations $|\psi_s|$ increases because of the rise of surfactant adsorption $\Gamma_1$. At higher concentrations the ionic strength becomes high enough, so that the effect of Debye screening prevails and leads to decrease of $|\psi_s|$.

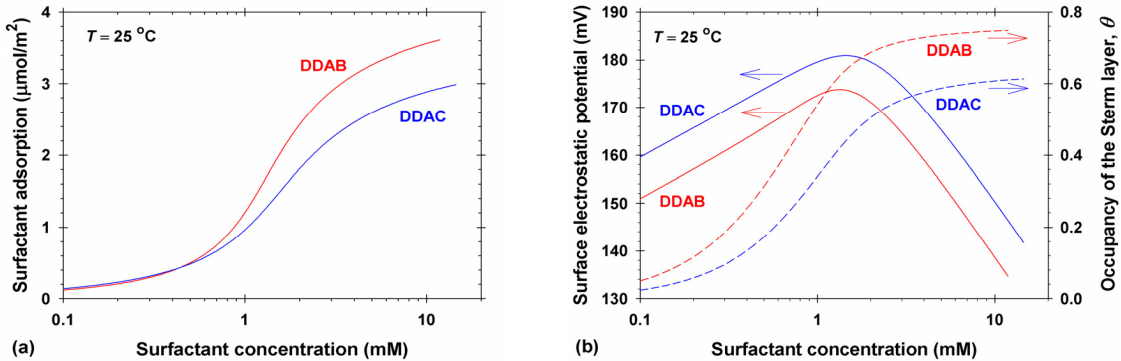

**Fig. E2**. Properties of the DDAB and DDAC adsorption layers calculated from the best data fits: (a) Surfactant adsorption, $\Gamma_1$, vs. concentration; (b) Surface electrostatic potential magnitude, $|\psi_s|$, and occupancy of the Stern layer, $\theta = \Gamma_2/\Gamma_1$, vs. surfactant concentration.

## Appendix F. Numerical solution of Poisson equation at concentrations near the CMC

In this case, we consider spherical micelles. The micellar concentration is low, so that we can solve the problem for a single micelle in infinite solution ($R_0 \to \infty$). The electrostatic boundary value problem is simpler than that in the case of the cell model (with finite $R_0$), because, at infinity the concentrations, $c_{1,0}$, $c_{2,0}$, and $c_{3,0}$, are known; at that $c_{2,0} = c_{1,0} + c_{3,0}$. For example, for SDS at the CMC (without added salt), we have $c_{1,0} = c_{2,0} = 8.08$ mM and $c_{3,0} = 0$; see Table C1. In general, the procedure is applicable to any concentrations of added salt. For this reason, we will work with the general expressions for the activity coefficients $\gamma_i$ – see Appendix D above.

The input parameters are $c_{1,0}$, $c_{2,0}$, $c_{3,0}$, $\Gamma_1$, $R_{el}$, $\lambda_B$ and $K_{St}$. For spherical geometry, the Poisson equation, Eq. (C5), reads:

$$\frac{d^2\Psi}{dr^2} + \frac{2}{r}\frac{d\Psi}{dr} = 4\pi\lambda_B(c_2 - c_1 - c_3) \quad \text{for } r > R_{el} \tag{F1}$$

where $R_{el}$ is the radius of the surface of charges. The boundary conditions at infinity are

$$\Psi \to 0 \text{ and } \frac{d\Psi}{dr} \to 0 \quad \text{for } r \to \infty \tag{F2}$$

The boundary condition at surface of the micelle reads:

$$\frac{d\Psi}{dr} = -4\pi\lambda_B(\Gamma_1 - \Gamma_2) \quad \text{for } r = R_{el} \tag{F3}$$

A generalization of the numerical approach in Ref. [34] is used. First, the following notations are introduced:

$$I_0 = c_{2,0}, \quad \kappa_0^2 = 8\pi\lambda_B I_0, \quad x \equiv \kappa_0 r \tag{F4}$$

where $I_0$ is the background ionic strength; $\kappa_0$ is the respective Debye parameter and $x$ is a dimensionless radial coordinate. Eqs. (F1) and (F3) acquire the form:

$$\frac{d^2\Psi}{dx^2} + \frac{2}{x}\frac{d\Psi}{dx} = \frac{c_2 - c_1 - c_3}{2I_0} \quad \text{for } x > x_{el} \equiv \kappa_0 R_{el} \tag{F5}$$

$$\frac{d\Psi}{dx} = -\frac{4\pi\lambda_B}{\kappa_0}\Gamma_1(1 - \frac{\Gamma_2}{\Gamma_1}) \quad \text{for } x = x_{el} \tag{F6}$$

At a large distance from the micelle, $x \gg x_{el}$, its electric field tends asymptotically to the field of a point charge:

$$\Psi = A\frac{\exp(-x)}{x} \Rightarrow \frac{d\Psi}{dx} = -(1+\frac{1}{x})A\frac{\exp(-x)}{x} = -(1+\frac{1}{x})\Psi \tag{F7}$$

where $A$ is a constant. For the needs of the computational procedure, we define a large enough distance, $x_0$, at which the solution for $\Psi(x)$ obeys Eq. (F7), viz.

$$\frac{d\Psi}{dx} + (1+\frac{1}{x})\Psi = 0 \quad \text{for } x = x_0 \tag{F8}$$

In our calculations we used $x_0 = 10$ for which $\exp(-x_0)/x_0 = 4.53999 \times 10^{-6}$. This leads to a relative error of $10^{-12}$ for the boundary condition, Eq. (F8). Note, that the boundary condition, Eq. (F6), combined with the Stern equation:

$$\frac{\Gamma_2}{\Gamma_1 - \Gamma_2} = K_{St}\gamma_{2,0}c_{2,0}\exp\Psi \quad \text{for } x = x_{el} \tag{F9}$$

gives

$$\frac{d\Psi}{dx} = -\frac{4\pi\lambda_B\Gamma_1}{\kappa_0}\frac{1}{1 + K_{St}\gamma_{2,0}c_{2,0}\exp\Psi} \quad \text{for } x = x_{el} \tag{F10}$$

The problem is solved by iteration procedure. To start the iterations, we use Eq. (F7) to define the zeroth-order order approximation for the electric potential:

$$\Psi_*(x) = \Psi_*(x_{el})\frac{x_{el}}{x}\exp(x_{el} - x) \tag{F11}$$

where $\Psi_*(x_{el})$ is determined from Eq. (F7) and the boundary condition, Eq. (F10). That is:

$$(1 + \frac{1}{x_{el}})\Psi_*(x_{el}) = \frac{4\pi\lambda_B\Gamma_1}{\kappa_0}\frac{1}{1 + K_{St}\gamma_{2,0}c_{2,0}\exp[\Psi_*(x_{el})]} \tag{F12}$$

For each $x$, the functions $c_1(\Psi_*(x))$, $c_2(\Psi_*(x))$ and $c_3(\Psi_*(x))$ are determined by solving numerically the system of three equations that originates from Eq. (C11):

$$\ln(\gamma_1 c_1) = \ln(\gamma_{1,0}c_{1,0}) - \Psi_*, \quad \ln(\gamma_2 c_2) = \ln(\gamma_{2,0}c_{2,0}) + \Psi_*, \quad \ln(\gamma_3 c_3) = \ln(\gamma_{3,0}c_{3,0}) - \Psi_* \tag{F13a}$$

where the activity coefficients, $\gamma_i = \gamma_i(c_1,c_2,c_3)$, $i = 1,2,3$, are determined by Eqs. (D3), (D6), (D7) and (D10), with parameter values given in Table D2. In the special case $\gamma_i \equiv 1$, Eqs. (F13a) are transformed into explicit expressions for $c_1(\Psi_*(x))$, $c_2(\Psi_*(x))$ and $c_3(\Psi_*(x))$ (Boltzmann equations):

$$c_1 = c_{1,0}\exp(-\Psi_*), \quad c_2 = c_{2,0}\exp(\Psi_*), \quad c_3 = c_{3,0}\exp(-\Psi_*) \tag{F13b}$$

The non-linear boundary value problem is represented in its linearization form for the new approximation, $\Psi_1$. From Eqs. (F5) one obtains:

$$\frac{d^2\Psi_1}{dx^2} + \frac{2}{x}\frac{d\Psi_1}{dx} = \frac{1}{2I_0}[c_2(\Psi_*) - c_1(\Psi_*) - c_3(\Psi_*)]$$

$$+\frac{(\Psi_1-\Psi_*)}{2I_0}\left(\frac{dc_2}{d\Psi}-\frac{dc_1}{d\Psi}-\frac{dc_3}{d\Psi}\right)_{\Psi=\Psi_*} \quad \text{for } x > x_{el} \tag{F14}$$

and the linearized form of the boundary conditions, Eqs. (F8) and (F10), reads:

$$\frac{d\Psi_1}{dx} = -\frac{4\pi\lambda_B\Gamma_1}{\kappa_0}\frac{1}{1+K_{St}\gamma_{2,0}c_{2,0}\exp\Psi_*}$$

$$+\frac{4\pi\lambda_B\Gamma_1}{\kappa_0}\frac{K_{St}\gamma_{2,0}c_{2,0}\exp\Psi_*}{(1+K_{St}\gamma_{2,0}c_{2,0}\exp\Psi_*)^2}(\Psi_1-\Psi_*) \quad \text{for } x = x_{el} \tag{F15}$$

$$\frac{d\Psi_1}{dx}+\left(1+\frac{1}{x}\right)\Psi_1 = 0 \quad \text{for } x = x_0 \tag{F16}$$

The linear problem, Eqs. (F14)–(F16), is solved numerically with the second order precision in the space. With the obtained new solution, $\Psi_1$, we calculate $c_1(\Psi_1)$, $c_2(\Psi_1)$, and $c_3(\Psi_1)$, and so on and so forth. Usually, ten iterations are enough to calculate the final solution of the considered problem with an excellent precision.

## References (Supplementary Information)